\documentclass[12pt]{article}
\usepackage{latexsym,epsfig,amssymb}
\usepackage{amsmath}
\usepackage{slashed}
\usepackage{mathrsfs}
\usepackage[bookmarks=false,colorlinks=true,linkcolor=blue,citecolor=blue,linktoc=page]{hyperref}
\newcommand{\w}{{\scriptstyle{\wedge}}\,}
\newcommand{\DIV}{C^{(4)}}

\newcommand{\ft}[2]{{\textstyle\frac{#1}{#2}}}

\newcommand{\eqn}[1]{(\ref{#1})}
\def\bfone{\relax{\rm 1\kern-.35em 1}}

\newcommand{\be}{\begin{equation}}
\newcommand{\ee}{\end{equation}}
\newcommand{\ben}{\begin{displaymath}}
\newcommand{\een}{\end{displaymath}}
\newcommand{\bea}{\begin{eqnarray}}
\newcommand{\eea}{\end{eqnarray}}

\newcommand{\bean}{\begin{eqnarray*}}
\newcommand{\eean}{\end{eqnarray*}}
\newcommand{\beqs}{\begin{eqnarray}}
\newcommand{\eeqs}{\end{eqnarray}}

\newcommand{\ca}[1]{{\mathcal{#1}}}
\renewcommand{\th}{h}
\newcommand{\tg}{g}

\newcommand{\mathon}{\mathversion{bold}}
\newcommand{\mathoff}{\mathversion{normal}}

\makeatletter
\@addtoreset{equation}{section}
\makeatother

\begin{document}

\thispagestyle{empty}

\begin{flushright}\small
MFA-11-36, ENSL-00625359
\end{flushright}

\bigskip
\bigskip

\mathon
\vskip 10mm
\begin{center}
  {\LARGE {\bf (1,0) superconformal models}\\[.5ex]{\bf in six dimensions}}
\end{center}
\mathoff


\vskip 4mm

\begin{center}
{\bf Henning Samtleben$^a$, Ergin Sezgin$^b$, Robert Wimmer$^a$}\\[3ex]

$^a$\,{\em Universit\'e de Lyon, Laboratoire de Physique, UMR 5672, CNRS et ENS de Lyon,\\
46 all\'ee d'Italie, F-69364 Lyon CEDEX 07, France} \\
\vskip 4mm

$^b$\,{\em George P. and Cynthia W. Mitchell Institute \\for Fundamental
Physics and Astronomy \\
Texas A\&M University, College Station, TX 77843-4242, USA}\\
\end{center}

\vskip1.8cm
\begin{center} {\bf Abstract } \end{center}
\begin{quotation}\noindent
We construct six-dimensional (1,0) superconformal models with non-abelian gauge
couplings for multiple tensor multiplets.
A crucial ingredient in the construction is the introduction of three-form gauge potentials
which communicate degrees of freedom between the tensor multiplets and the Yang-Mills
multiplet, but do not introduce additional degrees of freedom.
Generically these models provide only equations of motions. For a subclass
also a Lagrangian formulation exists, however it appears to exhibit
indefinite metrics in the kinetic sector.
We discuss several examples and analyze the
excitation spectra in their supersymmetric vacua.
In general, the models are perturbatively defined only  in the spontaneously broken
phase with the vev of the tensor multiplet scalars serving as the inverse
coupling constants of the Yang-Mills multiplet.
We briefly discuss the inclusion of hypermultiplets which complete the field content
to that of superconformal (2,0) theories.

\end{quotation}

\newpage
\setcounter{page}{1}

\tableofcontents
\newpage


\section{Introduction}


One of the discoveries of the seminal analysis in  \cite{Witten:1995zh}
is the existence of interacting quantum field theories in five and
six dimensions. Of particular interest are six-dimensional $(2,0)$
superconformal theories which are supposed to describe the low energy limit of
multiple coincident M5 branes.

However, no Lagrangian description for these theories is known and it is
in general believed that no such formulation exists: The M/string theory
origin implies that these theories have no free (dimensionless) parameter,
which would allow a parametrization to weak coupling and thus make the existence
of a Lagrangian description plausible. This conclusion was also drawn from
symmetry properties which imply that tree level amplitudes have to vanish \cite{Huang:2010rn}.
In addition, these $(2,0)$ theories
consist of chiral tensor multiplets and so far it has often been considered
as impossible to define non-abelian gauge couplings for such
multiplets.

 Regarding the first aspect the situation is similar to that of multiple M2 branes, as it was
before the recent developments that were triggered by the discovery of
the three dimensional $\ca{N}=8$ superconformal BLG model
\cite{Bagger:2007jr,Gustavsson:2007vu}. The meaning of this
$\ca{N}=8$ model in the M/string theory context is rather unclear, but
subsequently a $\ca{N}=6$ superconformal theory (ABJM model) was formulated
for an arbitrary number of M2 branes \cite{Aharony:2008ug}. The decisive
observation in \cite{Aharony:2008ug} is that an orbifold compactification
of the M theory/supergravity background provides a dimensionless,
though discrete, parameter $k$ which allows a parametrization to weak coupling
and thus also a Lagrangian formulation. The orbifold compactification
breaks $\ca{N}=8$ supersymmetry down to $\ca{N}=6$ except for $k=1,2$, where the
theory is strongly coupled. The $\ca{N}=6$ ABJM model has the same field content as the
$\ca{N}=8$ multiplet and it has been argued that monopole operators enhance
the supersymmetry to  $\ca{N}=8$ for $k=1,2$ \cite{Bashkirov:2010kz,Samtleben:2010eu}
(for $U(2)$ gauge group see \cite{Gustavsson:2009pm,Kwon:2009ar}).

We take here an analogous route. Instead of focusing on $(2,0)$ supersymmetry
we construct $(1,0)$ superconformal models for interacting multiple tensor multiplets.
One major obstacle, the nonabelian gauging of the (self dual) tensor fields,
is resolved by the introduction of a
tensor hierarchy \cite{deWit:2005hv,deWit:2008ta,Bergshoeff:2009ph}
which besides the Yang-Mills gauge field and the two-form gauge
potentials of the tensor multiplets contains also three-form gauge potentials.
 We therefore have an extended
tensor gauge freedom with $p=0,1,2$ $p$-form gauge parameters.

We then formulate essentially unique supersymmetry transformations for the various fields,
where we find a suitable extension of the structures introduced in \cite{Bergshoeff:1996qm}.
While in \cite{Bergshoeff:1996qm} the 2-form potential is a singlet, here it
carries a representation of the local gauge group, which is facilitated by the
introduction  of a 3-form potential that mediates couplings between the tensor and
vector multiplets. While the brane interpretation of our models requires further
investigation, it is worth mentioning that the field content of the model in  \cite{Bergshoeff:1996qm}
is known to arise in the worldvolume description of D6 branes stretch between
NS fivebranes \cite{Brunner:1997gk,Brunner:1997gf,Hanany:1997gh,Brunner:1997yq,Ferrara:1998vf}.
The closure of the supersymmetry algebra into translations and extended tensor gauge
transformations puts the system on-shell with a particular set of e.o.m.
For example the tensor multiplet field strength has to satisfy its self-duality condition
and the Yang-Mills field strength is related to the field strength of the three-form potentials
by a first-order duality equation.
Consequently, the three-form gauge potentials do not introduce additional degrees of freedom.
They communicate degrees of freedom between the
tensor multiplets and the Yang-Mills multiplet.
We also describe the extension of the tensor hierarchy
to higher p-form gauge potentials and briefly discuss the inclusion of hypermultiplets
which complete the field content to that of superconformal $(2,0)$ theories.

Consistency of the tensor hierarchy imposes a number of conditions on the possible gauge
groups and representations. We discuss several solutions of these conditions. Generically
these models provide only equations of motions, but for a subclass also a Lagrangian
formulation exists. In particular we find a Lagrangian model with $SO(5)$ gauge symmetry.
However, the existence of a Lagrangian description necessarily implies
indefinite metrics for the kinetic terms. It is at the moment not clear if the resulting
ghost states can be decoupled with the help of the large extended tensor gauge symmetry.
This and other questions regarding the quantization of the theory we have to leave for a
further investigation. A general feature of all considered cases is that the models
are perturbatively defined only in the spontaneously broken phase with the
vev of the tensor multiplet scalars serving as inverse coupling constants of the
Yang-Mills multiplets.

To write down a Lagrangian for a self dual field strength is in general a
formidable task. For a single M5 brane, in which case the e.o.m.\ are known
\cite{Howe:1997fb}, this has been done in~\cite{Pasti:1997gx,Bandos:1997ui,Cederwall:1997gg,Sezgin:1998tm}. We
consider these difficulties to be of a different category than finding a superconformal
non-abelian theory. When we formulate a Lagrangian we understand that the
first order duality equations are consistently imposed in addition to the second order
Lagrangian e.o.m., just as in the democratic formulation of ten-dimensional supergravity
\cite{Bergshoeff:2001pv}.

Finally we want to comment on some recent attempts and proposals for the description
of the $(2,0)$ theory. The low energy description of the theory when compactified
on a small circle is expected to be given by the maximal supersymmetric Yang-Mills
theory in five dimensions. Recent attempts tried to basically rewrite
five-dimensional Yang-Mills theory in six dimensions
\cite{Lambert:2010wm,Singh:2011id} or introduced non-abelian gaugings at
the cost of locality  \cite{Ho:2011ni}. Furthermore, it was recently proposed that
the $(2,0)$ theory is identical to five-dimensional super Yang-Mills theory for
arbitrary coupling
or compactification radius \cite{Douglas:2010iu,Lambert:2010iw}.
It is not clear yet
how one could obtain Yang-Mills theories in five dimensions
from the models presented here (even when including hypermultiplets).
Clearly a mechanism more complicated than a trivial dimensional reduction
has to be considered.

The paper is organized as follows: in section~\ref{sec:nonab} we present the general non-abelian
hierarchy of $p$-forms in six dimensions. We show that all couplings are parametrized in terms of a set
of dimensionless tensors that need to satisfy a number of algebraic consistency constraints.
In particular, we find that non-abelian charged tensor fields require the introduction of
St\"uckelberg-type couplings among the $p$-forms of different degree.
In section~\ref{sec:susyeom}, we extend the non-abelian vector/tensor system to a supersymmetric
system. Closure of the supersymmetry algebra puts the system on-shell and we derive the
modified field equations for the vector and tensor multiplets. In particular, we obtain the first-order
duality equation relating vector fields and three-form gauge potentials.
In section~\ref{subsec:hypers} we sketch the extension of the model upon inclusion of
hypermultiplets and gauging of their triholomorphic isometries.
In section~\ref{subsec:vacua} we derive the general conditions for maximally supersymmetric
vacua and compute the fluctuation equations by linearizing the equations of motion around such a vacuum.
Finally, we give in section~\ref{subsec:example} an explicit example with an arbitrary
compact gauge group and tensor fields transforming in the adjoint representation.

Section~\ref{sec:action} presents the additional conditions on the couplings in order to allow
for a Lagrangian formulation. We give the full action in section~\ref{subsec:action}.
In section~\ref{subsec:ex} we calculate the fluctuation equations induced by the action
and show that the degrees of freedom arrange in the free vector and self-dual tensor multiplet
as well as in certain `non-decomposable' combinations of the two.
We illustrate the general analysis in sections~\ref{subsec:solutions1} and ~\ref{subsec:solutions2}
with two explicit models that provide solutions to the consistency constraints
with compact gauge group $SO(5)$ and nilpotent gauge group $N_8$, respectively.
Finally, we summarize our findings in section~\ref{sec:conclusions}.


\section{Non-abelian tensor fields in six dimensions}
\label{sec:nonab}


In this section, we present the general (non-abelian) couplings of
vectors and anti-symmetric $p$-form fields in six dimensions.
While the standard field content of the ungauged theories falls into
vector and tensor multiplets, it is a general feature of these theories
that the introduction of gauge charges generically requires the
introduction of and couplings to three-form potentials.
The specific couplings can be derived successively and in a systematic way
by building up the non-abelian $p$-form tensor hierarchy, as
worked out in~\cite{deWit:2005hv,deWit:2008ta,Bergshoeff:2009ph},
see also \cite{Hartong:2009vc,Bergshoeff:2007ef,Gunaydin:2010fi} for some
applications to the specific $6D$ context.
Rather than going again step by step through the derivation of the general
couplings, we directly present the final result as
parametrized by a set of constant tensors (generalized structure constants)
that need to satisfy a system of algebraic consistency equations
(generalized Jacobi identities).
 In section~\ref{subsec:minimalh} we present the couplings for the minimal field content
 required to introduce non-abelian couplings between vector and tensor fields.
 In section~\ref{subsec:minimalh}, we extend the system to include also
 four-form gauge potentials.

\subsection{Minimal tensor hierarchy}
\label{subsec:minimalh}

The basic $p$-form field content of the theories to be discussed
is a set of vector fields~$A_\mu^r$, and two-form gauge potentials $B_{\mu\nu}^I$,
that we label by indices $r$ and $I$, respectively.
In addition, we will have to introduce three-form gauge potentials that we denote
by~$C_{\mu\nu\rho\,r}$.
The fact that three-form potentials are labeled by an index $r$ dual to the vector fields
is in anticipation of their dynamics: in six dimensions, these fields will be the on-shell
duals to the vector fields. For the purpose of this section however,
the dynamics of these fields is not yet constrained, the construction of the tensor
hierarchy remains entirely off-shell, and the indices `\,${^r}$\,' and `${_r}$' might be taken
as unrelated. Similarly, throughout this section, the self-duality of the field strength
of the two-form gauge potentials, which is a key feature of the later
six-dimensional dynamics, is not yet an issue.

The full non-abelian field strengths of vector and two-form gauge potentials are given as
\bea
{\cal F}_{\mu\nu}^r &\equiv&
2 \partial_{[\mu} A_{\nu]}^r - f_{st}{}^r A_\mu^s A_\nu^t + \th^r_I\,B_{\mu\nu}^I
\;,\nonumber\\[.5ex]
{\cal H}_{\mu\nu\rho}^I &\equiv& 3 D_{[\mu} B_{\nu\rho]}^I +
6 \, d^I_{rs}  A_{[\mu}^r \partial^{\vphantom{r}}_\nu A_{\rho]}^s
- 2 f_{pq}{}^s d^I_{rs} A_{[\mu}^r A_\nu^p A_{\rho]}^q
+ \tg^{Ir} C_{\mu\nu\rho\,r}
\;,
\label{defF}
\eea
in terms of the antisymmetric structure constants $f_{st}{}^r=f_{[st]}{}^r$,
a symmetric $d$-symbol $d^I_{rs}=d^I_{(rs)}$, and the tensors $g^{Ir}$, $h_I^r$
inducing St\"uckelberg-type couplings among forms of different degree.\footnote{
We use canonical dimensions such that a $p$-form has mass dimension $p$
and as a result all constant tensors $f_{st}{}^r$,
$d^I_{rs}$, $g^{Ir}$, $h_I^r$, are dimensionless.
}
Covariant derivatives are defined as $D_\mu \equiv \partial_\mu - A_\mu^r X_r$ with
an action of the gauge generators $X_r$ on the different fields
given by $X_r \cdot \Lambda^s \equiv - (X_{r})_{t}{}^s \Lambda^t$,
$X_r \cdot \Lambda^I \equiv - (X_{r})_{J}{}^I \Lambda^J$, etc.
The field strengths are defined such that they transform covariantly under the
set of non-abelian gauge transformations
\bea
\delta A_\mu^r &=& D_\mu \Lambda^r - \th^r_I \Lambda_\mu^I
\;,\nonumber\\[.5ex]
\Delta B_{\mu\nu}^I &=& 2 D_{[\mu} \Lambda_{\nu]}^I -2\, d^I_{rs} \Lambda^r {\cal F}_{\mu\nu}^s
- \tg^{Ir} \Lambda_{\mu\nu\,r}
\;,\nonumber\\[.5ex]
\Delta C_{\mu\nu\rho\,r} &=& 3 D_{[\mu} \Lambda_{\nu\rho]\,r}
+3 \, b_{Irs}\,{\cal F}_{[\mu\nu}^s \,\Lambda_{\rho]}^I
+ b_{Irs}\,{\cal H}_{\mu\nu\rho}^I \,\Lambda^s
+\dots
\;,\label{gaugesym}
\eea
where we have introduced the compact notation
\bea
\Delta B^I_{\mu\nu} &\equiv& \delta B^I_{\mu\nu} - 2 d^I_{rs}\,A_{[\mu}^r \,\delta A_{\nu]}^s
\;,\nonumber\\
\Delta C_{\mu\nu\rho\,r} &\equiv& \delta  C_{\mu\nu\rho\,r}
-3\, b_{Irs}\,B_{[\mu\nu}^I \,\delta A_{\rho]}^s
-2\, b_{Irs}\, d^I_{pq}\,A_{[\mu}^s \,A_{\nu}^p \,\delta A_{\rho]}^q
\;.\label{Delta1}
\eea
The ellipsis in the last line of (\ref{gaugesym}) represent possible terms that
vanish under projection with $\tg^{Ir}$.
This system is completely defined by the choice of the
invariant tensors $g^{Ir}$, $h_I^r$, $b_{Irs}$, $d^I_{rs}$, and $f_{rs}{}^t$.
It is obvious from (\ref{gaugesym}) that the shift symmetry
action on the $p$-form gauge fields can be used to gauge away
some of the $p$-forms, giving mass to others by the St\"uckelberg mechanism.
However, for the general analysis of couplings, we find it the most convenient
to work with the uniform system (\ref{gaugesym}) and to postpone possible gauge fixing
to the analysis of particular models\footnote{For a possible embedding into the general framework of $L_{\infty}$-algebra connections, we refer to \cite{Sati:2008eg}.
We thank U. Schreiber for pointing out this link.}.

Consistency of the tensor hierarchy requires that the gauge group generators
in the various representations are parametrized as
\bea
(X_{r})_{s}{}^t &=& -f_{rs}{}^t + d^I_{rs}\,\th_I^t
\;,\nonumber\\
(X_{r})_{I}{}^J &=&  2\,\th_I^s d^J_{rs}-\tg^{Js} b_{Isr}
\;,
\label{genpar}
\eea
in terms of the constant tensors appearing in the system.
The second relation exposes an important feature of the tensor hierarchy:
tensor fields can be charged under the gauge group only if either $\th_I^r$ or $\tg^{Ir}$ are non-vanishing,
i.e.\ they require some non-vanishing St\"uckelberg-type couplings in the field strengths (\ref{defF}).
This corresponds to the known results~\cite{Bekaert:1999dp,Bekaert:2000qx} that in absence of such couplings
and the inclusion of additional three-form gauge potentials, the free system of self-dual
tensor multiplets does not admit any non-abelian deformations.
On the other hand, the first relation of (\ref{genpar}) shows that in presence of $\th_I^r$, the
gauge group generators in the `adjoint representation' $(X_{r})_{s}{}^t$
are not just given by the structure constants but acquire a modification,
symmetric in its indices $(rs)$.

Furthermore, consistency of the system, i.e.\ covariant transformation behavior of the
field strengths (\ref{defF}) under the gauge transformations (\ref{gaugesym}) requires
the constant tensors to satisfy a number of algebraic consistency constraints.
A first set of constraints, linear in
$f$, $g$, $h$, is given by
\bea
2\left(d_{r(u}^J  d^I_{v)s}
-  d^I_{rs} d_{uv}^J \right) \th_J^s
&=& 2f_{r(u}{}^s d^I_{v)s} -b_{Jsr}  d_{uv}^J \,\tg^{Is}
\;,\nonumber\\[.5ex]
\left(d^J_{rs}\, b_{Iut}
+d^J_{rt}\, b_{Isu}
+2\, d^K_{ru} b_{Kst}\delta_I^J\right)\th_J^u
&=& f_{rs}{}^u b_{Iut}+f_{rt}{}^u b_{Isu}+\tg^{Ju} b_{Iur}  b_{Jst}
\;,
\label{lincon}
\eea
and ensures the invariance of the $d$- and the $b$-symbol under
gauge transformations.
The remaining constraints
are bilinear in $f$, $g$, $h$ and take the form
\bea
f_{[pq}{}^u f_{r]u}{}^s - \ft13 \th^s_I\, d^{I}_{u[p} f_{qr]}{}^u &=& 0
\;,\nonumber\\[.4ex]
{}
\th_I^r \tg^{Is} &=& 0
\;,\nonumber\\[.4ex]
f_{rs}{}^t \th_I^r - d^J_{rs}\,\th_J^t \th_I^r &=& 0
\;,\nonumber\\[.4ex]
{}
\tg^{Js} \th_K^r b_{Isr} - 2\th_I^s \th_K^r \, d^J_{rs} &=& 0
\;,\nonumber\\[.4ex]
{}
-f_{rt}{}^s g^{It} + d^J_{rt} h_J^s g^{It} - g^{It} g^{Js} b_{Jtr} &=& 0
\;.
\label{quadcon}
\eea
They may be understood as generalized Jacobi identities of the system:
together with (\ref{lincon}) they ensure the closure of the gauge algebra
according to
\bea
[X_r, X_s] &=& -(X_{r})_{s}{}^t \,X_t
\;,
\eea
for the generators (\ref{genpar}), as well as gauge invariance of the tensors $f$, $g$ and $h$.
The first equation of (\ref{quadcon}) shows that the standard Jacobi identity
is modified in presence of a non-vanishing $h^I_r$.
Even though the set of constraints (\ref{lincon}), (\ref{quadcon}) looks
highly restrictive, it admits rather non-trivial solutions and we will discuss explicit examples
of solutions
in sections~\ref{subsec:example},~\ref{subsec:solutions1}, and~\ref{subsec:solutions2}, below.
The system admits different abelian limits with $f_{rs}{}^t=0=g^{Ir}$ and
either $h_I^r$ or $d^{I}_{rs}$ vanishing, in which
the constraints (\ref{lincon}), (\ref{quadcon}) are trivially satisfied.
A slightly more general solution is given by vanishing $\th_I^r=0=\tg^{Ir}$
with $f_{rs}{}^t$ representing the structure constants of a Lie algebra.
With the particular choice $d^I_{rs}=d^I \delta_{rs}$, the vector-tensor
system then reduces to the coupling of the Yang-Mills multiplet to an uncharged
self-dual tensor multiplet as described in~\cite{Bergshoeff:1996qm}.

The covariant field strengths (\ref{defF}) satisfy the modified
Bianchi identities
\bea
D^{\vphantom{r}}_{[\mu} {\cal F}_{\nu\rho]}^r &=& \ft13 \th^r_I\,{\cal H}_{\mu\nu\rho}^I
\;,\nonumber\\[.5ex]
D^{\vphantom{I}}_{[\mu} {\cal H}_{\nu\rho\sigma]}^I &=&
\ft32 d^I_{rs} \, {\cal F}^r_{[\mu\nu} {\cal F}^s_{\rho\sigma]}+
\ft14 \tg^{Ir} {\cal H}^{(4)}_{\mu\nu\rho\sigma\,r}
\;,
\label{Bianchi}
\eea
where the non-abelian field strength ${\cal H}^{(4)}_{\mu\nu\rho\sigma\,r}$
of the three-form potential is defined by the second equation.
In turn, its Bianchi identity is obtained from (\ref{Bianchi}) as
\bea
D^{\vphantom{(}}_{[\mu} {\cal H}^{(4)}_{\nu\rho\sigma\tau]\, r} &=&
-2\,b_{I rs}\,{\cal F}_{[\mu\nu}^s\,{\cal H}_{\rho\sigma\tau]}^I
+ \dots
\;,
\label{Bianchi2}
\eea
where the ellipsis represents possible terms that
vanish under projection with $\tg^{Ir}$.
We finally note that the general variation of the field-strengths is given by
\bea
\delta {\cal F}_{\mu\nu}^r &=& 2 D_{[\mu} \delta A_{\nu]}^r + \th^r_I\,\Delta B_{\mu\nu}^I
\;,\nonumber\\[.5ex]
\delta {\cal H}_{\mu\nu\rho}^I &=& 3 D_{[\mu} \Delta B_{\nu\rho]}^I
+6\, d^I_{rs} \, {\cal F}_{[\mu\nu}^r \,\delta A_{\rho]}^s + \tg^{Ir}\,\Delta C_{\mu\nu\rho\,r}
\;,\nonumber\\[.5ex]
\delta {\cal H}^{(4)}_{\mu\nu\rho\sigma\,r} &=& 4 D_{[\mu} \Delta C_{\nu\rho\sigma]r}
-6\, b_{Irs}\,{\cal F}_{[\mu\nu}^s\,\Delta B_{\rho\sigma]}^I
+ 4\, b_{Irs}\,{\cal H}_{[\mu\nu\rho}^I\,\delta A_{\sigma]}^s + \dots
\;,\label{Delta2}
\eea
again with the ellipsis representing possible terms that
vanish under projection with~$\tg^{Ir}$.

\subsection{Extended tensor hierarchy}
\label{subsec:extendedh}

The field content introduced in the last section were the $p$-forms
$A_\mu^r$, $B_{\mu\nu}^I$, $C_{\mu\nu\rho\,r}$,
for which in particular we have defined their non-abelian field strengths.
Strictly speaking, in the entire system, only a subset of the three-form potentials
have appeared, defined by projection with the tensor~$\tg^{Ir}$ as $g^{Ir} C_{\mu\nu\rho\,r}$\,.
As it turns out, this truncation is precisely the `minimal field content' required
in order to write down an action and/or define a consistent set of equations of motion.
Off-shell on the other hand, the tensor hierarchy may be extended to the full
set of three-form potentials, which then necessitates the introduction
of four-form gauge potentials, etc.

For later use, we present in this section the results of the general tensor hierarchy for the four-form gauge
potentials which we denote by $\DIV_{\mu\nu\rho\lambda\,\alpha}$ with covariant
field strength ${\cal H}^{(5)}_{\alpha}$\,. The full version of the Bianchi identity (\ref{Bianchi2}) then reads
\bea
D^{\vphantom{(}}_{[\mu} {\cal H}^{(4)}_{\nu\rho\sigma\tau]\, r} &=&
-2\,b_{I rs}\,{\cal F}_{[\mu\nu}^s\,{\cal H}_{\rho\sigma\tau]}^I
+
\ft15 k_{r}{}^{\alpha} \, {\cal H}^{(5)}_{\mu\nu\rho\sigma\tau\,\alpha}
\;,
\label{Bianchi3}
\eea
where now the field strength ${\cal H}^{(5)}_{\alpha}$ itself satisfies the Bianchi identity
\bea
D^{\vphantom{(}}_{[\mu} {\cal H}^{(5)}_{\nu\rho\lambda\sigma\tau]\,\alpha}&=&
\ft{10}{3}c_{\alpha\,IJ} {\cal H}^I_{[\mu\nu\rho} {\cal H}^J_{\lambda\sigma\tau]}
-\ft{5}{2}c^{t}_{\alpha\,s} {\cal F}^s_{[\mu\nu} {\cal H}^{(4)}_{\rho\lambda\sigma\tau]\,t}
~+\cdots
\;,
\label{Bianchi4}
\eea
up to terms vanishing under projection with the tensor $k_{r}{}^{\alpha}$.
The new constant tensors $k_r{}^\alpha$, $c_{\alpha\,IJ}$, and $c^{t}_{\alpha\,s}$
are constrained by the relations
\bea
k_{r}{}^{\alpha} c_{\alpha\,IJ} &=& \th_{[I}^s b_{J]rs}\;, \quad
k_{r}{}^{\alpha} c^{t}_{\alpha\,s} ~=~ f_{rs}{}^t-b_{Irs} \tg^{It}+d^I_{rs} \th_I^t
\;,\quad
\tg^{Kr}\,k_{r}{}^{\alpha}~=~0
\;,
\label{conk}
\eea
which extend the constraints (\ref{lincon}), (\ref{quadcon}).
As a consistency check,
we note that equations (\ref{lincon}), (\ref{quadcon}) imply the
orthogonality relations
\bea
\tg^{Kr}\, h_{[I}^s b^{\vphantom{s}}_{J]rs} &=& 0
\;,\nonumber\\[.5ex]
\tg^{Kr}\, \left( f_{rs}{}^t - g^{It}b_{Irs}+h^t_I d^I_{rs} \right) &=& 0
\;,
\label{ortho}
\eea
showing that (\ref{conk})
does not imply new constraints among the previous tensors.
Furthermore, consistency of the extended system requires an additional
relation among $b$- and $d$-symbol to be satisfied
\bea
b^{\vphantom{J}}_{Jr(s}  d_{uv)}^J   &=& 0
\;,
\label{mag}
\eea
as also noted in~\cite{Hartong:2009vc}.
The new tensor gauge transformations take the form
\bea
\Delta C_{\mu\nu\rho\,r} &=& 3 D_{[\mu} \Lambda_{\nu\rho]\,r}
+3 \, b_{Irs}\,{\cal F}_{[\mu\nu}^s \,\Lambda_{\rho]}^I
+ b_{Irs}\,{\cal H}_{\mu\nu\rho}^I \,\Lambda^s
-k_{r}{}^{\alpha}\,\Lambda_{\mu\nu\rho\,\alpha}
\;,
\nonumber\\[.5ex]
\Delta \DIV_{\mu\nu\rho\sigma\,\alpha} &=& 4\,D_{[\mu} \Lambda_{\nu\rho\sigma]\,\alpha}
-8\,c_{\alpha\,IJ}\, {\cal H}_{[\mu\nu\rho}^{[I}\,\Lambda_{\sigma]}^{J]}
+6\,c^{t}_{\alpha\,s}\, {\cal F}_{[\mu\nu}^s\,\Lambda_{\rho\sigma]\,t}
\nonumber\\
&&{}
+c^{t}_{\alpha\,s}\,{\cal H}^{(4)}_{\mu\nu\rho\sigma\,t}\,\Lambda^s
~+\dots
\;,\qquad
\eea
where the first equation completes the corresponding transformation law
of (\ref{gaugesym}) and the second transformation is given up to
terms that vanish under projection with the tensor $k_{r}{}^{\alpha}$.
Accordingly, the general variation of the non-abelian field strengths
from (\ref{Bianchi3}), (\ref{Bianchi4}) is given by
\bea
\delta {\cal H}^{(4)}_{\mu\nu\rho\sigma\,r} &=& 4 D_{[\mu} \Delta C_{\nu\rho\sigma]r}
-6\, b_{Irs}\,{\cal F}_{[\mu\nu}^s\,\Delta B_{\rho\sigma]}^I
+ 4\, b_{Irs}\,{\cal H}_{[\mu\nu\rho}^I\,\delta A_{\sigma]}^s +
k_{r}{}^{\alpha}\,\Delta \DIV_{\mu\nu\rho\sigma\,\alpha}
\;,
\nonumber\\[1ex]
\delta{\cal H}^{(5)}_{\mu\nu\rho\sigma\tau\,\alpha}
&=&5\,D_{[\mu} \Delta \DIV_{\nu\rho\sigma\tau]\,\alpha}
-10\,c^{t}_{\alpha\,s}\, {\cal F}_{\mu\nu}^s \, \Delta C_{\rho\sigma\tau]\,t}
-20 c_{\alpha\,IJ}\,{\cal H}^{[I}_{[\mu\nu\rho}\,\Delta B^{J]}_{\sigma\tau]}
\nonumber\\
&&
{}
-5\,c^{t}_{\alpha\,s}\,\delta A_{[\mu}^s\,{\cal H}^{(4)}_{\nu\rho\sigma\tau]\, t}
~+\dots
\;.
\label{deltaH5}
\eea
Continuing along the same line, the tensor hierarchy can be continued by introducing
five-form and six-form potentials together with their field strengths and non-abelian
gauge transformations. For the purpose of this paper we will only need the vector/tensor
system up to the four-form gauge potentials given above.


\section{Superconformal field equations}
\label{sec:susyeom}


In the previous section we have introduced the tensor hierarchy for
$p$-form gauge potentials ($p=1,2,3$) with the associated
generalized field strengths (\ref{defF}) and Bianchi identities (\ref{Bianchi}).
Gauge covariance w.r.t.\ the extended tensor gauge symmetry (\ref{gaugesym})  implies
a number of conditions on the (dimensionless) invariant tensors and generators of the gauge group
(\ref{genpar})--(\ref{quadcon}), but otherwise does not contain any information about the
dynamics of theses fields.

The aim of this section is to complete the bosonic fields of the tensor hierarchy into supersymmetry
multiplets in order to obtain a non-abelian superconformal model for
the $(1,0)$ vector and tensor multiplets. With the given (bosonic) field content of the
tensor hierarchy (\ref{defF}), a supersymmetric tensor hierarchy will contain Yang-Mills
multiplets $(A^r_\mu, \lambda^{i\,r}, Y^{ij\,r})$, and tensor multiplets
$(\phi^I, \chi^{i\, I}, B_{\mu\nu}^I)$, labeled by indices $r$ and $I$, respectively.
The index $i=1,2$ indicates the $Sp(1)$ R-symmetry, the field $Y^{ij}$ denotes the
auxiliary field of the off-shell vector multiplets.
In addition one has to accommodate within this structure
the three-form potential $C_{\mu\nu\rho\, r}$ whose presence was crucial in the last section
 in order to describe non-abelian charged tensor fields.

\subsection{Supersymmetry}

The coupling of a single  $(1,0)$ self-dual tensor multiplet to a Yang-Mills multiplet was introduced
in \cite{Bergshoeff:1996qm} and as a first step we give the necessary generalization for
a non-abelian coupling of an arbitrary number of these tensor multiplets.
To this end, we introduce supersymmetry transformations such that they close into
translations and the extended tensor gauge symmetry (\ref{gaugesym}) according to
\bea{}
[\delta_{\epsilon_1}, \delta_{\epsilon_2} ] &=&
\xi^\mu \partial_\mu + \delta_\Lambda + \delta_{\Lambda_\mu} +  \delta_{\Lambda_{\mu\nu}}
\;,
\label{close}
\eea
with field dependent transformation parameters for the respective transformations. These parameters
are given by
\bea
\xi^\mu &\equiv& \ft12\bar{\epsilon_2} \gamma^\mu \epsilon_1
\;,\nonumber\\
\Lambda^r &=& -\xi^\mu A_\mu^r
\;,\nonumber\\
\Lambda^I_\mu &=& -\xi^\nu B_{\nu\mu}^I+d^I_{rs} \Lambda^r A_\mu^s  + \xi_\mu\,\phi^I
\;,\nonumber\\
\Lambda_{\mu\nu\,r} &=& -\xi^\rho C_{\rho\mu\nu\,r}-b_{Irs}\Lambda^s B_{\mu\nu}^I
-\ft23 b_{Irp}d^I_{qs} \Lambda^s A_{[\mu}^p A_{\nu]}^q
\;,\label{para}
\eea
as will be shown shortly.
With $d^I_{rs}=\alpha' d^I \delta_{rs}$, $b_{Irs}=0$, this reproduces the corresponding algebra of~\cite{Bergshoeff:1996qm}.\footnote{Note that in canonical dimensions,
the tensor $d^I_{rs}$ is dimensionless.}
The supersymmetry transformations for the Yang-Mills multiplet are given by
\bea
\delta A^r_\mu&=& -\bar\epsilon\gamma_\mu\lambda^r\;,\nonumber\\[.5ex]
\delta \lambda^{i\,r} &=& \ft18\,\gamma^{\mu\nu} {\cal F}^r_{\mu\nu} \epsilon^i
-\ft12\,Y^{ij\,r}\epsilon_j + \ft14 \th^r_I \phi^I \epsilon^i \;,\nonumber\\[.5ex]
\delta Y^{ij\,r} &=& - {\bar\epsilon}^{(i}\gamma^\mu D_\mu\lambda^{j)r}
+2 \th^r_I\,\bar\epsilon^{(i} \chi^{j) I}
\;.
\label{YMsusy}
\eea
Here the generalization w.r.t.\ the transformations for the off-shell pure Yang-Mills multiplet
is parametrized by the constant tensor $\th_I^r$ and brings in the fields
$(\phi^I, \chi^{i\, I}, B_{\mu\nu}^I)$ of the tensor multiplets on the r.h.s.\ of the transformations.
These additional terms are necessary for the
supersymmetry algebra to close to the generalized tensor gauge symmetry (\ref{close}), (\ref{para}).
E.g.\ the last term in $\delta \lambda^{i\,r}$ is required to produce the proper $\delta_{\Lambda_\mu}$
action in the commutator of supersymmetries on the vector field $A^r_\mu$.
Likewise, the last term in $\delta Y^{ij\,r}$ ensures the proper closure of the supersymmetry algebra
on $\lambda^{i\,r}$. It then comes as a non-trivial consistency check, that the variation of this last term
is precisely what is needed for closure of the algebra on $Y^{ij\,r}$.
Even though, fields from the tensor multiplets appear in these transformation rules,
the Yang-Mills multiplet by itself, using the necessary tensor multiplet transformations,
still closes off-shell.

Next we give
the supersymmetry transformations of the tensor multiplet
\bea
\delta \phi^I &=& \bar\epsilon\chi^I\;,\nonumber\\
\delta\chi^{i\,I} &=& \ft1{48} \, \gamma^{\mu\nu\rho} \,{\cal H}^{I}_{\mu\nu\rho}
\epsilon^i +\ft14\,\gamma^\mu D_\mu\phi^I \epsilon^i
- \ft12 d^I_{rs} \gamma^\mu\lambda^{i\,r}\,
\bar\epsilon\gamma_\mu \lambda^s\;,
\nonumber\\
\Delta B^I_{\mu\nu} &=& -\bar\epsilon\gamma_{\mu\nu}\chi^I\;,
\nonumber\\[1ex]
\Delta C_{\mu\nu\rho\,r} &=& - b_{Irs} \, \bar\epsilon\gamma_{\mu\nu\rho}\lambda^s \phi^I
\;,\label{tensusy}
\eea
where we have used the same notation (\ref{Delta1}) for general variation
introduced in the tensor hierarchy. We also note that
$\gamma^{\mu\nu\rho}\epsilon^i$ acts as a self-duality projector such that only
${\cal H}^{I\,+}_{\mu\nu\rho}$, see (\ref{Hplus}), is actually alive in $\delta\chi^{i\,I}$.
W.r.t.~the couplings discussed in~\cite{Bergshoeff:1996qm},
the r.h.s.\ of these transformations has been generalized
by the introduction of the general $d$-symbol,
and the inclusion of covariant field strengths and derivatives on the now charged fields of the tensor multiplets.
In particular, the important new ingredient in these transformation rules is the three-form potential
$C_{\mu\nu\rho\,r}$ which is contained in the definition of $\ca{H}^I_{\mu\nu\rho}$
and contributing to its supersymmetry transformation according to (\ref{Delta2}).
Its presence has been vital in establishing the non-abelian bosonic vector-tensor
system in the last section, and similarly, its presence turns out to be indispensable for
closure of the supersymmetry algebra here.
To group it with the tensor multiplet in (\ref{tensusy})
is a mere matter of convenience; with the same right it might be
considered as a member of the gauge multiplet
(indeed, as mentioned before by its dynamics the three-form potential will be
the dual of the vector fields $A_\mu^r$).
The form of its supersymmetry transformation (\ref{tensusy}),
mixing Yang-Mills and tensor multiplet fields, displays its dual role as a messenger between
these two multiplets.
Note that we have given in (\ref{tensusy})
the supersymmetry transformation for the uncontracted three-form $C_{\mu\nu\rho\,r}$,
although all the explicit couplings only contain the contracted expression  $g^{K\,r}C_{\mu\nu\rho\,r}$.
We will come back to this difference in the following.

Closure of the supersymmetry algebra on the tensor multiplet according to (\ref{close}) is now rather
non-trivial and heavily relies on the extra terms arising from variation of the three-form potential.
In particular, the algebra closes only on-shell on the tensor multiplets. In the search for new model or theory such a property may be considered as feature that provides a certain uniqueness.
We will discuss these equations and their individual origin now in detail.

\subsection{Minimal model}
\label{subsec:minimalm}

We first investigate the equations of motion resulting from
supersymmetrization of the bosonic field content of the minimal tensor hierarchy
of section~\ref{subsec:minimalh}. In particular, this model includes only the projected subset
$g^{K\,r} C_{\mu\nu\rho\,r}$ of three-form gauge potentials.
The resulting tensor multiplet field equations are given by
\bea
{\cal H}^{I\,-}_{\mu\nu\rho} &=&-  d^I_{rs} \bar\lambda^r \gamma_{\mu\nu\rho} \lambda^s
\;,
\nonumber\\[1ex]
\gamma^\sigma D_\sigma \chi^{iI} &=&
\ft12  d^I_{rs} {\cal F}^r_{\sigma\tau}\, \gamma^{\sigma\tau}\lambda^{is}
+  2 d^I_{rs} Y^{ij\,r}\, \lambda^{s}_j
+\left(d^I_{rs}\th^s_J - 2 b_{J sr}\tg^{Is}  \right) \phi^J \lambda^{ir}
\;,
\nonumber\\[1ex]
D^\mu D_\mu\,\phi^I &=&
-\ft12d^I_{rs} \left(
 {\cal F}_{\mu\nu}^r {\cal F}^{\mu\nu\, s} -4\,Y_{ij}^{r} Y^{ij\,s} + 8 \bar\lambda^r \gamma^\mu D_\mu \lambda^s
\right)
\nonumber\\[.4ex]
&&{}
-2 \left(b_{J sr}\tg^{Is} -8 d^I_{rs}\th^s_J  \right) \bar\lambda^r \chi^J
-3\, d^I_{rs}\th_J^r\th^s_K\,
\phi^J\phi^K
\;.
\label{eomten}
\eea
The first equation, which imposes a self duality condition on the three-from field strength,
originates in the closure of supersymmetry on the associated two-form potential $B_{\mu\nu}^I$.
The closure on $\delta\chi^{i\,I}$ gives the fermionic equations of motion while
the scalar field equation is obtained by the supersymmetry transformation of the
$\chi^{iI}$- equation. The fact that the tensor fields are charged under the gauge group
has rather non-trivial
consequences, namely supersymmetry variation of the field equations (\ref{eomten})
in turn implies the following first-order equations of motion for the Yang-Mills multiplets
\bea
\tg^{Kr}b_{Irs}\,\left( Y^s_{ij} \,\phi^I - 2 \bar\lambda^s_{(i} \chi_{j)}^I \right) &=& 0
\;,
\nonumber\\[1ex]
\tg^{Kr}b_{Irs}\left({\cal F}_{\mu\nu}^s \phi^I-2\,\bar\lambda^s \gamma_{\mu\nu} \chi^I \right)
&=& \ft{1}{4!}\,\varepsilon_{\mu\nu\lambda\rho\sigma\tau}\,\tg^{Kr}\,{\cal H}^{(4)\,\lambda\rho\sigma\tau}_{r}
\;,
\nonumber\\[1ex]
\tg^{Kr}b_{Irs} \left(
\phi^I \gamma^\mu D_{\mu} \lambda_i^{s}
+\ft12 \gamma^\mu  \lambda_i^{s} D_{\mu}\phi^I
\right) &=&
\tg^{Kr}b_{Irs} \left(
\ft14  {\cal F}_{\mu\nu}^s\gamma^{\mu\nu}  \chi_i^{I}
+\ft1{24} {\cal H}_{\mu\nu\rho}^I\gamma^{\mu\nu\rho}  \lambda_i^{s}
- Y_{ij}^s\,\chi^{j\,I}
\right.
\nonumber\\[.5ex]
&&{}
\qquad\quad\;\;\left.{}
+ \ft32 \th^s_J \phi^I \chi_i^{J}
+\ft13 d^I_{uv}\,\gamma^\mu \lambda_i^{u} \bar\lambda^s \gamma_\mu \lambda^v
\right)
\;.
\nonumber\\
\label{eomYM}
\eea
The first equation is the algebraic equation for the auxiliary field $Y^{ij\,r}$, while the second equation
provides the anticipated duality of vector fields and three-form potentials by relating their respective
field strengths. In particular, derivation of this equation and use of the Bianchi identity (\ref{Bianchi2})
gives rise to a standard second-order equation of Yang-Mills type for the vector fields $A_\mu^r$\,.
Equivalently, the first two equations of (\ref{eomYM}) can be inferred from closure
of the supersymmetry algebra on the three-form gauge potentials
$\tg^{Kr} C_{\mu\nu\rho\,r}$\,.
The appearance of the Yang-Mills dynamics (\ref{eomYM}) from supersymmetry of
the tensor field equations (\ref{eomten}) is in strong contrast to the model
of~\cite{Bergshoeff:1996qm} (in which effectively $\tg^{Kr}=0$, and the tensor
field are not charged) where the vector fields remain
entirely off-shell or can alternatively be set on-shell with field equations that do not contain the
tensor multiplet fields.
Moreover, in the model of~\cite{Bergshoeff:1996qm}, an algebraic equation analogous to
the first equation of (\ref{eomYM}) is excluded by the appearance of an anomaly in its
supersymmetry variation (see also~\cite{Howe:1998ts}).
We should stress that in the present model, such anomalies are actually absent
due to the particular Fierz identities (\ref{id1}), (\ref{id2})
in combination with the identity (\ref{mag}). I.e.\ the cubic fermion terms in the
supersymmetry variation of (\ref{eomYM}) cancel precisely, which yields a strong consistency check
of the construction.

To summarize, the system of equations of motion (\ref{eomten}), (\ref{eomYM})
consistently transforms into itself under supersymmetry.
It describes a novel system of supersymmetric non-abelian couplings for
multiple $(1,0)$ tensor multiplets in six dimensions. The equations of motion contain no
dimensionful parameter and hence the system is at least classically (super)-conformal.
A crucial ingredient to the model are the three-form gauge potentials $C_{\mu\nu\rho\,r}$
which are related by first-order duality equations to the vector fields of the theory
and thus do not constitute new dynamical degrees of freedom.
This is similar
to the situation of Chern-Simons matter theories in the context of multiple $M2$ branes
\cite{Aharony:2008ug,Bagger:2007jr}.
The actual model depends on the explicit
choice of the gauge group and representations and the associated invariant tensors of
the gauge group which have to satisfy the conditions (\ref{genpar})--(\ref{quadcon}).
The task that remains is to find explicit solutions for these constraints. We will discuss different
examples in sections 3.6, 4.4 and 4.5 below.

\subsection{Extended model}
\label{subsec:extm}

The above described model represents the minimal field content and equations of motion,
required for closure of the supersymmetry algebra and the supersymmetry of the equations of motions.
In particular, it relies on the projected subset
$g^{K\,r} C_{\mu\nu\rho\,r}$ of three-form gauge potentials.
Just as for the bosonic tensor hierarchy in section~\ref{subsec:extendedh},
one may seek to extend the above supersymmetric
system to the full set of three-form gauge potentials.
With the supersymmetry transformation of general
$C_{\mu\nu\rho\,r}$ given by (\ref{tensusy}),
closure of the supersymmetry algebra leads to the following uncontracted equations
\bea
b_{Irs}\,\left( Y^s_{ij} \,\phi^I - 2 \bar\lambda^s_{(i} \chi_{j)}^I \right) &=& 0
\;,
\nonumber\\[1ex]
b_{Irs}\left({\cal F}_{\mu\nu}^s \phi^I-2\,\bar\lambda^s \gamma_{\mu\nu} \chi^I \right)
&=& \ft{1}{4!}\,\varepsilon_{\mu\nu\lambda\rho\sigma\tau}\,{\cal H}^{(4)\,\lambda\rho\sigma\tau}_{r}
\;,
\nonumber\\[1ex]
b_{Irs} \left(
\phi^I \gamma^\mu D_{\mu} \lambda_i^{s}
+\ft12 \gamma^\mu  \lambda_i^{s} D_{\mu}\phi^I
\right) &=&
b_{Irs} \left(
\ft14  {\cal F}_{\mu\nu}^s\gamma^{\mu\nu}  \chi_i^{I}
+\ft1{24} {\cal H}_{\mu\nu\rho}^I\gamma^{\mu\nu\rho}  \lambda_i^{s}
- Y_{ij}^s\,\chi^{j\,I}  \;+
\right.
\nonumber\\[.5ex]
&&{}
\quad\;\;\left.{}
+  \th^s_J \left(2\phi^I \chi_i^{J}-\ft12\phi^J \chi_i^{I}\right)
+\ft13 d^I_{uv}\,\gamma^\mu \lambda_i^{u} \bar\lambda^s \gamma_\mu \lambda^v
\right)
\;,
\nonumber\\
\label{eomYM_uncontracted}
\eea
In order to have this system close under supersymmetry it is necessary to introduce also
a four-form gauge potential.
Consequently the tensor hierarchy has to be continued one step further
as described in section~\ref{subsec:extendedh}.
The resulting supersymmetry transformation of the four-form potential is
\bea
\Delta \DIV_{\mu\nu\rho\sigma\,\alpha} &=&
2c_{\alpha\,IJ}\,\phi^{[I}\,\bar\epsilon\gamma_{\mu\nu\rho\sigma} \chi^{J]}\;,
\eea
Furthermore, supersymmetry of the field equations (\ref{eomYM_uncontracted}) induces the
first-order field equations
\bea
\ft1{5!}\,\varepsilon_{\mu\nu\rho\lambda\sigma\tau}\,k_{r}{}^{\alpha}\,{\cal H}^{(5)\,\mu\nu\rho\lambda\sigma}_{\alpha}
 =
2 k_{r}{}^{\alpha}\left( c_{\alpha\,IJ}\left(\phi^I D_\mu \phi^J-2\bar\chi^I \gamma_\mu \chi^J \right)
-c_{\alpha\,u}^{t}b_{Jtv}\,\bar\lambda^u\gamma_\mu \lambda^v\right)
\;.
\label{H5}
\eea
This shows that the dynamics of $\DIV_{\mu\nu\rho\sigma\,\alpha}$ is given by
a first-order duality equations, which relates these four-form potentials
to the Noether current of some underlying global symmetry. In particular, this
first-order equation ensures
that the four-form gauge potentials do not constitute new dynamical degrees of freedom.

\subsection{Adding hypermultiplets}
\label{subsec:hypers}

Another possible extension of the supersymmetric model presented above is
the inclusion of hypermultiplets\footnote{For superconformal couplings of hypermultiplets to Yang-Mills, see \cite{Ivanov:2005kz}.}. As is well known, global supersymmetry requires the hyperscalars
to parametrize a hyper-K\"ahler manifold ${\cal M}_{\rm h}$,
more precisely superconformal symmetry requires ${\cal M}_{\rm h}$ to be a hyper-K\"ahler cone.
The above presented non-abelian theories can be extended to include gaugings
of isometries on the hyper-K\"ahler cone along the lines
of~\cite{Sierra:1983uh,deWit:2001bk,deWit:1999fp},
from which the additional couplings and in particular the resulting scalar potential
can be inferred.
While we defer the details of this extension to another publication, here
we only sketch a few relevant elements of the construction.
Within in the above construction, gauging of triholomorphic isometries on the hyper-K\"ahler cone
is achieved by introducing an embedding tensor $\vartheta_r{}^\alpha$
that encodes
the coupling of vector fields $A^r_\mu$ to hyper-K\"ahler isometries $K_\alpha$ and is subject to the
algebraic conditions
\bea
f_{pr}{}^s \vartheta_s{}^\alpha &=&
f_{\beta\gamma}{}^\alpha \vartheta_{p}{}^\beta  \vartheta_r{}^\gamma\;,
\qquad
h^r_I\,\vartheta_r{}^\alpha ~=~ 0
\;,
\eea
with the structure constants $f_{\alpha\beta}{}^\gamma$ of the algebra of
hyper-K\"ahler isometries.
On the other hand, in the presence of hypermultiplets, the vector multiplet equations of motion
(\ref{eomYM_uncontracted}) allow for a consistent modification,
in particular in the $Y$-field equation as
\bea
b_{Irs} \left( Y^s_{ij} \,\phi^I - 2 \bar\lambda^s_{(i} \chi_{j)}^I \right) &=&
k_r{}^\alpha P_\alpha^{ij}
\;,
\eea
with the constant tensor $k_r{}^\alpha$ from (\ref{conk}), and the moment
maps $P_\alpha^{ij}$ associated with the triholomorphic hyper-K\"ahler isometries.
It is only by means of this algebraic equation for $Y^s_{ij}$ that the hyperscalars
enter the tensor multiplet field equations.
Further requiring the existence of an action eventually leads to
the identification
\bea
k_r{}^\alpha &=& \vartheta_r{}^\alpha
\;,
\eea
i.e.\ relates the gauging of hyper-K\"ahler isometries to a modification of the
vector and tensor multiplet field equations.

\subsection{Supersymmetric vacua and excitation spectrum}
\label{subsec:vacua}

We study now supersymmetric vacua for the minimal model of section~\ref{subsec:minimalm}
and the excitation spectrum in such vacua, i.e.\ the linearized field equations.
The algebraic equation for the vector field strength, the second equation
in (\ref{eomYM}), indicates that the expectation value of the tensor multiplet scalar $\phi^I$
serves as an (inverse) coupling constant. This notion will become more evident in the subsequent
sections where we discuss models which provide a Lagrangian. Consequently, the perturbative
analysis is limited to the spontaneously broken phase where $\phi^I$ has a (large) expectation value.

The Killing spinor equations of the theory (\ref{action}) are obtained from (\ref{YMsusy}), (\ref{tensusy})
\bea
0&\stackrel{!}\equiv& \delta \lambda^{i\,r} ~=~  \ft18\,\gamma^{\mu\nu} {\cal F}^r_{\mu\nu} \epsilon^i
-\ft12\,Y^{ij\,r}\epsilon_j + \ft14 \th^r_I \phi^I \epsilon^i \;,\nonumber\\[1ex]
0&\stackrel{!}\equiv& \delta\chi^{i\,I} ~=~ \ft1{48} \, \gamma^{\mu\nu\rho} {\cal H}^{I\,+}_{\mu\nu\rho}
\epsilon^i +\ft14\,\gamma^\mu D_\mu\phi^I \epsilon^i
\;,
\label{KS}
\eea
and characterize solutions that preserve some fraction of supersymmetry.
These equations show that a Lorentz-invariant solution preserving all supersymmetries
corresponds to setting the scalar fields to constant values $\phi^I_0$ satisfying
\bea
\phi^I_0 \,\th_{I}^r &=& 0
\;,
\label{vacuum0}
\eea
and setting all other fields to zero. Expanding the scalar fluctuations as $\phi^I \equiv \phi^I_0 +\varphi^I$
and imposing the condition (\ref{vacuum0}) one obtains
at the linearized level for the field equations (\ref{eomten}), (\ref{eomYM}) the system:
\begin{align}
  \label{eomlin}
   (\,dB^I + g^{Ir} C_r\,)^{-} = 0\;,& \hspace{4.2cm}   N^I_r\, Y^r_{ij} = 0\;, \nonumber\\[3pt]
   \slashed{\partial}\, \chi^{iI} + 2\,N^I_r\, \lambda^{ir} = 0\;,& \hspace{2.15cm}
    N^I_r dA^r - g^{Ir}{} \,^\ast dC_r = 0\;,  \nonumber\\[3pt]
   \Box \varphi^I - N^I_r \, \partial \cdot A^r = 0\;,& \hspace{3.98cm}
    N^I_r \, \slashed{\partial} \lambda ^{ir} = 0\;,&
\end{align}
where we have defined the matrices
\bea
K_{rs}&\equiv& \phi_0^I \, b_{Irs}
\;,\qquad
N_r^I ~\equiv~ \tg^{Is}\, K_{sr}
\;.
\label{KN}
\eea
and used that $N^I_r\,h^r_J  = 0$, by the first
identity in  (\ref{ortho}) and the susy vacuum condition~(\ref{vacuum0}).

\paragraph{Unbroken gauge symmetry.} For a generic supersymmetric vacuum which satisfies
(\ref{vacuum0}) the vector gauge transformations $\Lambda^r\,X_r$
are broken down to the subgroup
of transformations $\Lambda^{\underline{r}}\,X_{\underline{r}}$
which satisfy
\begin{align}
  \label{unbrok}
   X_{\underline{r}\,J}{}^I\,\phi_0^J ~=~ -N^I_{\underline{r}} ~\overset{!}{=}~ 0\, ,
\end{align}
where the index $\underline{r}$ labels the subset of unbroken generators (\ref{genpar}).
The rest of the extended tensor gauge symmetry (\ref{gaugesym}) remains intact.
Consequently,
in the case that the gauge group is not completely broken, the matrix $N^I_r$, and for invertible
$g^{Ir}$ also the matrix $K_{rs}$, always has some null-directions. The fluctuation equations (\ref{eomlin})
show that for these null-directions the fields of the corresponding vector multiplets drop out of this perturbative
analysis. This is nothing else than the above mentioned observation that the perturbative analysis
is valid only in the spontaneously broken phase and that the unbroken sector of the Yang-Mills
multiplet is (infinitely) strongly coupled and perturbatively not visible. This part of the spectrum
decouples and should be integrated out for a proper treatment.

In general it is rather difficult to break the gauge symmetry completely with a single
scalar field. The addition of hypermultiplets as sketched in section~\ref{subsec:hypers}
may offer additional possibilities in this directions.
This is for example comparable with the situation of $\ca{N}=2$ SQCD, for which mixed Coulomb-Higgs phases
with vev's for vector multiplet and hypermultiplet scalars exist where the theory is
completely higgsed. In such a case there would be regions in the  moduli space of vacua
where the complete spectrum of the models discussed here is perturbatively accessible.
For the extended models of section~\ref{subsec:extm} on the other hand, the coupling of the
Yang-Mills multiplet is given by the matrix $K_{rs}$ which may have less null
directions than the matrix $N^I_r$.

\subsection{A model with adjoint tensor multiplets}
\label{subsec:example}

A particular solution to the constraints (\ref{lincon}), (\ref{quadcon})
is given by choosing some semi-simple compact gauge group $G$ with Lie-algebra $\mathfrak{g}$,
identifying both $I$ and $r$ with the adjoint representation of $G$, and
the tensor $\tg^{rs}$ with the Cartan-Killing metric. Moreover we set
\bea
h_r^s &\equiv& 0\;,\qquad
d^p_{rs} ~\equiv~ d_{rst} \tg^{pt}\;,\qquad
b_{p\,rs} ~\equiv~ f_{prs}\;,
\label{ex1}
\eea
with the totally symmetric $d$-symbol $d_{rst}$ and the
totally antisymmetric structure constants $f_{rst}$\,.
As will be discuss in detail in the next section, for a solution of this form
the resulting theory does not admit an action and is described by the set of
equations of motion
(\ref{eomten}), (\ref{eomYM}) only.

With $\tg^{rs}$ being the (invertible) Cartan-Killing metric, the matrices $N$ and $K$ introduced in
(\ref{KN}) are essentially the same,
\begin{equation}
  \label{KNr}
  N_s^r = g^{rt}\, K_{ts} =:K^r{}_s = -\phi_0^t\,f_{ts}{}^r \, ,
\end{equation}
and the matrix $K^r{}_s$ defines the adjoint action of the vev $\phi_0$. By a gauge
rotation the $\phi_0$ can always be chosen to lie in the Cartan subalgebra~$\mathfrak{t}$,
and we decompose $\mathfrak{g}$ into the orthogonal sum
$\mathfrak{g}=\mathfrak{t}\oplus \tilde{\mathfrak{g}}$\,.
In that case, the unbroken sector of the Yang-Mills multiplets,
which drops out of the fluctuation equations, spans
the Cartan subalgebra $\mathfrak{t}$ on which the action of $K^r{}_s$ vanishes.
On the orthogonal complement $\tilde{\mathfrak{g}}$ and for generic choice of $\phi_0$,
the matrix $K^r{}_s$ is invertible, and using the Cartan-Weyl basis we introduce the notation
$K_{\mathrm{red}}^{\ \ \tilde r}{}_{\, \tilde s} = k_{\tilde r}\,\delta^{\tilde r}_{\tilde s}$
for the reduced matrix on this subspace (there is no summation over repeated indices in this case).

Before giving the explicit excitation equations for this specific model we
discuss the gauge fixing of the vector field gauge symmetry, which for $h_r^s =0$ is an
ordinary gauge symmetry, see (\ref{gaugesym}). A convenient gauge, which
disentangles the scalar field and gauge field fluctuations is given by the
Lorenz gauge condition
\begin{equation}
  \label{lgauge}
    \partial\cdot A^r = 0\, .
\end{equation}
Since the gauge fields are determined by first-order equations the Lorenz gauge, and not a
't Hooft $R_\xi$-gauge, decouples the scalar and gauge field kinetic terms.
The fluctuation equations (\ref{eomlin}) thus take the form
\begin{align}
  \label{eomlinEx1}
    \slashed{\partial}\, \chi^{i\underline {r}} =0\;,& \hspace{3.32cm}  dC^{\,\underline{r}} = 0\;,&
    \nonumber\\[2pt]
   (\,dB^r + C^r\,)^{-} = 0\;,& \hspace{3.35cm}    \Box \varphi^r  = 0\;,&
   \nonumber\\[1pt]
   \slashed{\partial}\, \chi^{i\tilde r} + 2\,k_{\tilde r}\, \lambda^{i\tilde r} = 0\;,&
    \hspace{1.5cm}
     dA^{\tilde r} - \frac{1}{k_{\tilde r}}\, {}^\ast dC^{\tilde r} = 0\;,&
     \nonumber\\[1pt]
 \slashed{\partial} \lambda ^{i\tilde r} = 0\;,  & \hspace{3.65cm}
   Y^{\tilde r}_{ij} = 0\;,  &
\end{align}
where we have split the gauge indices as $r= (\underline{r}, \tilde r)$
according to the decomposition
$\mathfrak{g}=\mathfrak{t}\oplus \tilde{\mathfrak{g}}$.
For the unbroken sector $\mathfrak{t}$,
the first line of (\ref{eomlinEx1})
together with the second line for $r={\underline{r}}$  thus describe a
free tensor multiplet coupled to the three-form potential $C^{\underline{r}}$
which has vanishing field strength and may be gauged away.
Alternatively, one may employ the two-form shift symmetry in (\ref{gaugesym}) with gauge
parameter $\Lambda^{\underline{r}}_{\mu\nu}$  to set $B^{\underline{r}}=0$.
Then the linearized equations describe a self-dual closed field $C^{\underline{r}}$ which
gives an equivalent description of the free tensor multiplet.

The broken sector $\tilde{\mathfrak{g}}$
is described by the second line of (\ref{eomlinEx1}) for $r=\tilde{r}$ together with
the last two lines.
Here also the Yang-Mills
multiplet is present but the structure is somewhat unusual.  The multiplet structure
is not the direct sum of a free tensor and Yang-Mills multiplet, but forms a
multiplet that we call henceforth non-decomposable, as can be seen in particular from the fermionic
field equations. This seems to be a general feature of the models considered here and will
be discussed in section \ref{subsec:ex}.
The third equation in the right column again demonstrates the
dual role of the three-form potential $C^{\tilde{r}}$: Acting with $d\,^\ast$ on this equation
implies the second-order free field equation $\Box A^{\tilde r}=0$. The original equation then fixes
$C^{\tilde{r}}$ in terms of  $A^{\tilde r}$ up to an two-form $b^{\tilde r}$ whose field strength has to be
self dual in the  $B^r=0$ gauge, see the second line of (\ref{eomlinEx1}). The three-form
potential~$C^{\tilde{r}}$ therefore shifts or communicates degrees of freedom between the gauge and
tensor multiplet.

\paragraph{Note added:}
Another example for a tensor gauge symmetry that satisfies the constraints (\ref{lincon}), (\ref{quadcon}), 
without specifying any supersymmetric dynamics,
has recently been given in~\cite{Chu:2011fd}. In this example, the tensor fields $B_{\mu\nu}^r$ are in the adjoint representation of a compact gauge group $G$, and vector fields are taken to appear in two copies of the adjoint representation $A_\mu^r$, $A_\mu^{r'}$. The choice
\bea
f_{rs}{}^t &=&  \mathsf{f}_{rs}{}^t\;,\qquad f_{rs'}{}^{t'} ~=~ - f_{s'r}{}^{t'} ~=~ \ft12 \mathsf{f}_{rs'}{}^{t'}
\;,\nonumber\\
d^t_{rs'} &=& d^t_{s'r} ~=~ -\ft12 \mathsf{f}_{rs'}{}^t\;,
\qquad h_s^{r'} ~=~ \delta_s^{r'}
\;,
\label{chu}
\eea
with all other components vanishing, defines a solution of (\ref{lincon}), (\ref{quadcon}) in terms of the structure constants $\mathsf{f}_{rs}{}^t$ of the group $G$. The resulting model does not involve any coupling to three-forms ($g^{rs}=0$) but exhibits the St\"uckelberg-type coupling between vector fields $A_\mu^{r'}$ and the tensors. The supersymmetric field equations for the tensor multiplet are as given in (\ref{eomten}), with (\ref{chu}) used, while the Yang-Mills multiplet is off-shell. An action that produces the tensor multiplet field equations can be easily written down, with the self-duality condition on the 3-form field strength understood to be imposed after the variation of the action, but it can also easily be checked that such an action is not supersymmetric.

\section{Action}
\label{sec:action}

So far, we have found a set of field equations that consistently
transform into each other under $(1,0)$ supersymmetry.
The full system is entirely determined by the choice of the
constant tensors $g^{Ir}$, $h_I^r$, $b_{Irs}$, $d^I_{rs}$, and $f_{rs}{}^t$
subject to the set of algebraic constraints (\ref{lincon}), (\ref{quadcon}).
In this section we present the additional conditions, which these tensors
have to satisfy in order for the field equations to be integrated to an action.
We give the full supersymmetric action and discuss the general structure
of supersymmetric vacua and the fluctuation equations around such vacua.
The non-unitarity of the action manifests itself in the generic appearance
of some unusual `non-decomposable' multiplet couplings.
Finally, we illustrate the general analysis by two concrete models,
with compact gauge group $SO(5)$ and a nilpotent eight-dimensional gauge group, respectively.

\subsection{Conditions for existence of an action}

The existence of an action first of all requires the existence of a constant non-degenerate
metric $\eta_{IJ}$ by which tensor multiplet indices
can be raised and lowered, in order to provide a kinetic term for the scalar fields
and the other fields of the tensor multiplets.
Further inspection of the field equations (\ref{eomten})--(\ref{eomYM_uncontracted})
then shows that their integrability to an action
requires the identifications
\bea
\th_I^r = \eta_{IJ} \tg^{Jr}\;,\qquad d^I_{rs} = \ft12 \eta^{IJ} b_{Jrs}
\;,
\label{res_action}
\eea
i.e.\ in particular a $b$-symbol that is symmetric in its indices $(rs)$.
Moreover, in the process of computing the action, one finds that
the identity (\ref{mag}) needs to be imposed in order to ensure the existence
of a proper topological term.
From (\ref{res_action}) it is obvious that the models we have discussed
in section~\ref{subsec:example} indeed do not admit an action.

To summarize, with these identifications,
the algebraic consistency conditions (\ref{lincon}), (\ref{quadcon}), (\ref{mag}) reduce to
\bea
b_{I\,r(u}b^I_{vs)} &=& 0
\;,\nonumber\\[.4ex]
{}
\left(b^J_{r(u}\, b^I_{v)s}
-b^J_{uv}\,b^I_{rs}
+ b_{K\,rs} b^K_{uv}\, \eta^{IJ}\right)\tg_J^s
&=& 2f_{r(u}{}^s b^I_{v)s}
\;,\nonumber\\[.4ex]
6f_{[pq}{}^u f_{r]u}{}^s -\tg^s_I\, b^{I}_{u[p} f_{qr]}{}^u &=& 0
\;,\nonumber\\[.4ex]
{}
2f_{rs}{}^t \tg_I^r -  b^J_{rs}\,\tg_J^t \tg_I^r &=& 0
\;,\nonumber\\[.4ex]
g_K^r \tg_{[I}^{s}b^{\vphantom{s}}_{J]sr} &=& 0
\;,\nonumber\\[.4ex]
\tg_I^r \tg^{Is} &=& 0
\;.
\label{conaction}
\eea
Finding non-trivial solutions to these constraints is a formidable task. We will
give in sections~\ref{subsec:solutions1},~\ref{subsec:solutions2}
below some explicit solutions that are inspired from
similar constructions in gauged supergravity theories.

\subsection{The action}
\label{subsec:action}

In case the constant tensors satisfy all algebraic conditions (\ref{conaction}), the equations of motion
(\ref{eomten}), (\ref{eomYM}) can be lifted to an action.
In fact, one may verify a somewhat stronger conclusion: the identifications (\ref{res_action})
and thus the set of constraints (\ref{conaction})
appear already to be necessary in order to construct
a conserved supercurrent
underlying the equations of motion (\ref{eomten}), (\ref{eomYM})
from a canonical structure for the fermions~\cite{Burke:2011mw}.

In order to write an action,
we ignore for the moment the subtleties of writing an action for a self-dual three-form field strength,
but give a standard second-order action, keeping in mind that the corresponding first-order equation
of (\ref{eomten}) is supposed to be imposed after having derived the
second-order equations of motion, just as in the democratic formulation
of ten-dimensional supergravities~\cite{Bergshoeff:2001pv}.\footnote{
Alternatively, this self-duality can be implemented by using a non-abelian
version~\cite{Samtleben:2011eb} of the Henneaux-Teitelboim action~\cite{Henneaux:1988gg}
that breaks manifest space-time covariance.}
The full action then reads
\bea
{\cal L} &=&
-\ft18 D^\mu \phi_I \,D_\mu \phi^I
-\ft12 \bar\chi_I\, \gamma^\mu D_\mu \chi^{I}
+\ft1{16} b_{I rs} \phi^I \left(
 {\cal F}_{\mu\nu}^r {\cal F}^{\mu\nu\, s}
-4 Y_{ij}^{r} Y^{ij\,s} +8  \bar\lambda^r \gamma^\mu D_{\mu} \lambda^{s} \right)
\nonumber\\[.5ex]
&&{}
-\ft1{96}  {\cal H}_{\mu\nu\rho}^I\, {\cal H}^{\mu\nu\rho}_I
-\ft1{48} b_{Irs} {\cal H}_{\mu\nu\rho}^I\,\bar\lambda^r\gamma^{\mu\nu\rho}  \lambda^{s}
- \ft14 b_{Irs} {\cal F}_{\mu\nu}^r\,\bar\lambda^s\gamma^{\mu\nu}  \chi^{I}
+ b_{Irs} Y_{ij}^r\,\bar\lambda^{i\,s}\chi^{j\,I}
\nonumber\\[.4ex]
&&{}
+  \ft12\left( b_{J sr}\tg_I^{s}  -4  b_{I sr}\tg_J^{s} \right) \phi^I  \bar\lambda^r \chi^J
+  \ft18  b_{Irs}\tg_J^r\tg^s_K\,\phi^I \phi^J\phi^K
-\ft1{48} {\cal L}_{\rm top}
\nonumber\\[.5ex]
&&{}
-\ft1{24} b_{Irs} b^I_{uv}\,\bar\lambda^r\gamma^\mu \lambda^{u} \bar\lambda^s \gamma_\mu \lambda^v
\;,
\label{action}
\eea
which shows explicitly the role of the scalar fields $\phi^I$ as inverse coupling constants
for the Yang-Mills multiplet.
Like the equations of motion, this action contains no
dimensionful parameter such that the system is (super)-conformal at least at the classical level.
The topological term is given by integrating
\bea
dV\delta {\cal L}_{\rm top}
= 6\left\{
b_{I r s}\,\delta A^r\w {\cal F}^s\w {\cal H}^{I}
 -\Delta B^I\w\left( g_I^r\,{\cal H}^{(4)}
-\tfrac{1}{2}b_{Irs} {\cal F}^r\w{\cal F}^s\right)
- g_I^r \Delta C_{r}\w {\cal H}^{I}\right\}
\,,
\eea
and has the explicit form
\bea
dV{\cal L}_{\rm top} \!\!\!&=&\!\!\!
-6\, \tg_I^r\,  C_r \w {\cal H}^{I}
+  b_{Irs}  B^I\w {\cal F}^r \w {\cal F}^s
- b_{Irs}h_J^rh_K^s\,B^I\w B^J\w B^K
\nonumber\\[.5ex]
&&{}\hspace{-5mm}
 + B^I\w \left[
 h_I^s b^J_{su}b_{Jvr} A^u\w A^v\w d A^r
+\ft34 (b_{Irs}f_{pq}{}^r +4 b_{Jqs}X_{p\,I}{}^J)\,f_{uv}{}^sA^p\w A^q\w A^u\w A^v
\right]
\nonumber\\[.5ex]
&&{}\qquad
- \ft1{10}\,f_{up}{}^s b^J_{qs}b_{Jvr}\,
A^p\w A^q\w A^u\w A^v\w d A^r
\;.
\eea
It can be understood in compact form as the boundary contribution of a manifestly
gauge-invariant seven-dimensional term
\bea
\int_{\partial M_7} {\cal L}_{\rm top}   &\propto& \int_{M_7}
\left(b_{Irs}\, {\cal F}^r \w {\cal F}^s \w {\cal H}^I
-  {\cal H}^I \w { D} {\cal H}_I
\right)
\;.
\eea
As usual, gauge invariance of the topological term may lead to quantization conditions
for the various coupling constants.
For the tensor multiplet, it is straightforward to verify that
the action (\ref{action}) induces the field equations (\ref{eomten}) from above
with the first order equation imposed by hand.
For the fields of the vector multiplet, we obtain the first and the last of the uncontracted
equations (\ref{eomYM_uncontracted}), while the duality equation relating ${\cal F}_{\mu\nu}^r$ and
${\cal H}^{(4)}_{\mu\nu\rho\sigma\,r}$ only appears in its contracted form (\ref{eomYM}). In addition,
variation w.r.t.\ the vector field gives rise to the Yang-Mills equation
\bea
b_{Irs} \,D^\nu \left( \phi^I {\cal F}_{\mu\nu}^s - 2 \bar\lambda^s\gamma_{\mu\nu}\chi^I \right)
&=&
\left(\phi^ID_\mu \phi^J-2 \bar\chi^I \gamma_\mu \chi^J\right) X_{r\,IJ}
-2 \phi^I b_{I pq} X_{rs}{}^q\, \bar\lambda^p \gamma_\mu \lambda^s
\nonumber\\[.5ex]
&&{}
-\ft1{12}b_{Irs}\, \varepsilon_{\mu\nu\rho\lambda\sigma\tau}\,
{\cal F}^{\nu\rho\,s} {\cal H}^{\lambda\sigma\tau\,I}
\;,
\label{YM}
\eea
that can alternatively obtained as a derivative of the uncontracted duality equation (\ref{eomYM_uncontracted})
upon use of the first-order equation (\ref{H5}).

We note that the last constraint equation of (\ref{conaction}) shows that non-trivial solutions
to these constraints (i.e.\ solutions in which the tensor fields are charged)
exist only if the metric $\eta_{IJ}$ is indefinite,
which in turn implies that some of the scalars (and some of the two-forms) in (\ref{action})
have a negative kinetic term. This somewhat reminds the situation for the three-dimensional
BLG theories \cite{Bagger:2007jr,Gustavsson:2007vu}
with Lorentzian three-algebra~\cite{Gran:2008vi,Gomis:2008uv,Benvenuti:2008bt,Ho:2008ei,Bandres:2008kj},
and certainly requires further investigation.
We also note that similar structures as encountered in this section
have appeared in generic 6d supergravity theories
 \cite{Nishino:1986dc,Sagnotti:1992qw,Ferrara:1997gh,Ferrara:1996wv,Nishino:1997ff,Riccioni:2001bg}.

We conclude with a presentation of the superconformal symmetry transformations \cite{Sezgin:1994th}. Denoting the fields in the theory by $\Phi=(\phi^I, B_{\mu\nu}^I,\chi^I,A_\mu^r, Y^{ij},\lambda^r, C_{\mu\nu\rho r})$, the conformal transformations are given by
\be
\delta_C \Phi = {\cal L}_\xi \Phi  + \lambda_D \Omega \Phi\ ,
\ee
where ${\cal L}_\xi$ is the Lie derivative with respect to the conformal Killing vector defined by
$\partial_{(\mu} \xi_{\nu)}=\Omega\eta_{\mu\nu}$, which also defines $\Omega$,
and $\lambda_D$ are the Weyl weight for $\Phi$ given by $(2,0,5/2,0,2,3/2,0)$. The Lie derivative for the fermionic fields $\Psi=(\chi^I, \lambda^r)$, in particular, takes the form ${\cal L}_\xi \Psi = \xi^\mu\partial_\mu\Psi +\frac14 \partial_\mu\xi_\nu \gamma^{\mu\nu}\Psi$. The conformal supersymmetry transformations, on the other hand, involve conformal Killing spinors, consisting of a pair $(\eta_+, \eta_-)$ that satisfy $\partial_\mu\eta_+ -\frac12 \gamma_\mu\eta_-=0$. The superconformal transformations take the form of supersymmetry transformations in which the constant supersymmetry parameter $\epsilon$ is replaced by $\eta_+$, and the parameter $\eta_-$ arises only in $\delta_{\eta_-} \chi^I = -\frac12 \phi^I \eta_-$. Note that the bosonic conformal transformation together with supersymmetry ensures the full superconformal symmetry since the commutator of conformal boost with supersymmetry yields the special supersymmetry generator \cite{Bergshoeff:1985mz}.

\subsection{Multiplet structure of excitations}
\label{subsec:ex}

The supersymmetry transformations of the model (\ref{action})
are still given by equations~(\ref{YMsusy}), (\ref{tensusy}),
such that the Killing spinor equations remain of the form (\ref{KS}).
In particular, the existence of a maximally supersymmetric vacuum
is still encoded in the condition (\ref{vacuum0})
on the scalar expectation values.
In this vacuum, the linearized field equations obtained from (\ref{action})
extend the fluctuation equations (\ref{eomlin}) by the linearization of
the second-order equation for the vector fields (\ref{YM}),
which takes the form
\bea
 K_{rs} \left( \Box\,A_\mu^s-\partial_\mu \partial^\nu A_\nu^s
 \right) &=&
  N_{Ir} \left(\partial_\mu \varphi^I - \partial^\nu B_{\nu\mu}^I \right)
  \;.
\eea
With the gauge fixing $\partial^\nu B_{\nu\mu}^I=0=
\partial^\mu A_\mu^r+g^r_I \varphi^I$, the equation turn into the
free Klein-Gordon equation for the vector field components $A_\mu^r$.\footnote{
Alternatively, this can be achieved by choosing Lorenz gauge for the vector fields
and fixing the tensor gauge freedom by
$\partial^\nu B_{\nu\mu}^I\equiv\partial_\mu \varphi^I$.
This is a consistent gauge choice since the scalar field equation in this gauge turns
into the massless Klein-Gordon equation.}
With this gauge fixing, the full set of
linearized field equations obtained from (\ref{action}) is given as
\begin{align}
   (\,dB^I + g^{Ir} C_r\,)^{-} =&~ 0\;, \hspace{4.8cm}    K_{rs}\, Y^s_{ij} = 0\;, \nonumber\\[3pt]
   \slashed{\partial}\, \chi^{iI} + 2\,N^I_r\, \lambda^{ir} =& ~0\;, \hspace{2.8cm}
    N^I_r dA^r - g^{Ir}{} \,^\ast dC_r = 0\;,  \nonumber\\[3pt]
   \Box \varphi^I  =& ~0\;, \hspace{2.6cm}
    K_{rs} \, \slashed{\partial} \lambda ^{is} - 2 N_{rI}\chi^{iI} = 0\;,& \nonumber\\[3pt]
      K_{rs}\, \Box\,A_\mu^s =&~0
  \;,
\label{eom_linearized_action}
\end{align}
with the matrices $K_{rs}$ and $N^I_r$ from (\ref{KN}). We note that
$N^I_r g^{rJ} = 0 = g^{rI}N_I^s$\,.
With a proper choice of basis such that $K_{rs}$ is diagonal, the lowest
order dynamics contains $r_K = \mathrm{rank}(K)$ vector multiplets. The fluctuation
equations (\ref{eom_linearized_action}) decouple into various multiplets
which we denote as follows, and whose multiplicities are given in Table 1:
\bea
{\rm (V)} &:& \Box A_\mu = 0 \,,
\quad \slashed{\partial} \lambda = 0\,,
\nonumber\\[1ex]
{\rm (T)} &:& \Box \varphi = 0 \,,\quad \slashed{\partial} \chi = 0\,,\quad
(dB)^{-} = 0\,,
\nonumber\\[1ex]
{\rm (T')} &:& \Box \varphi = 0 \,,\quad \slashed{\partial} \chi = 0\,,\quad
(dB)^{-} = -g C^-\,,\quad
dC = 0 \,,
\nonumber\\[1ex]
{\rm (TV)} &:&
\Box \varphi = 0
\,,\quad
\kappa\, dA = \,^* \!dC
\,,\quad
(dB)^- = -g C^-
\,,\quad
\slashed{\partial} \lambda = 0
\,,\quad
\slashed{\partial} \chi = -2g\kappa\lambda
\,,
\nonumber\\[1ex]
{\rm (VT)} &:&
\Box \varphi = 0
\,,\quad
\Box A_\mu = 0 \,,
\quad
(dB)^- = 0
\,,\quad
\slashed{\partial} \chi = 0
\,,\quad
\slashed{\partial} \lambda  = 2g\, \chi
\,.
\label{excitations}
\eea
We have kept the coupling constants $g$ and $\kappa$ to keep track of the scales of
$g^{Ir}$ and $\phi_0^I$, respectively.
The first two multiplets (V) and (T) are the free vector and self-dual tensor multiplet, respectively,
the third one (T') is the self-dual tensor multiplet enhanced by a non-propagating three-form potential.
The fourth line (TV) describes the `non-decomposable' combination of a free vector multiplet and
a self-dual tensor multiplet for which the vector multiplet acts as a source. It is obvious from the fermionic
field equations that these two multiplets cannot be separated. This is the type of coupling we have encountered in
the broken sector $\tilde{\mathfrak{g}}$ of the model described in section~\ref{subsec:example}.
The last line (VT) describes the dual version of such a `non-decomposable' coupling, here a free self-dual
tensor multiplet acts as the source for a vector multiplet.
This situation is similar to the observation made in \cite{Howe:1998ts}
regarding the BSS model \cite{Bergshoeff:1996qm}.
Diagonalizing for example the $\chi$-equation and using the relations for $N^I_r$ given above
shows that there are $r_N= \mathrm{rank}(N)$ TV-multiplets.
Is straightforward to verify that only the combination of (TV) and (VT) can be derived from an action,
which implies that they appear with equal multiplicity and thus the $\lambda$-equation in
(\ref{eom_linearized_action}) describes $r_K-2\, r_N$ vector multiplet (VT) fermions.
In a similar fashion one finds the multiplicities of the other couplings in
(\ref{excitations}) as obtained from the equations (\ref{eom_linearized_action}) and which are
entirely encoded
in the rank of the matrices $g^{Ir}$, $N_r^I$ and $K_{rs}$. We collect the explicit result in
table~\ref{tab:mult}. In the following, we will illustrate these general structures in some
explicit examples.

\begin{table}[bt]
\begin{center}
\begin{tabular}{|c||c|c|c|c|c|}
\hline
multiplet &
(V) & (T) & (T') & (TV) & (VT)
\\ \hline
\# &
$r_K - 2 r_N$ &
$n_{\rm T} - r_N-r_g$ &
$r_g-r_N$ &
$r_N$ &
$r_N$ \\
\hline
\end{tabular}
\caption{Multiplicities of the different structures (\ref{excitations})
appearing in the lowest order fluctuations (\ref{eom_linearized_action})
expressed in terms of the number of tensor multiplets $n_{\rm T}$
and the ranks $r_g$, $r_N$, $r_K$ of the matrices $g^{Ir}$, $N_r^I$ and $K_{rs}$,
respectively.}
\label{tab:mult}
\end{center}
\end{table}

\subsection{Example: $SO(5)$ gauge group}
\label{subsec:solutions1}

The constraints (\ref{conaction}) constitute a rather non-trivial system of
consistency conditions for the undetermined constant tensors and structure constants.
Fortunately, a number of solutions can be inferred from analogous
construction in gauged supergravity theories.
In this section, we discuss a solution to~(\ref{conaction}) that is
inspired by gaugings of the maximal six-dimensional supergravity theory~\cite{Bergshoeff:2007ef}.

Let the indices $I$ and $r$ parametrize the vector and spinor
representations of the group $SO(5,5)$, respectively,
let $\eta_{IJ}$ to be corresponding invariant metric, and set
\bea
b^I_{rs}&\equiv&\gamma^I_{rs}
\;,\qquad
f_{rs}{}^t ~\equiv~
- 4\,\gamma^{IJK}_{rs}\gamma_{IJ\,p}{}^t\,\tg_K^p
\;,
\label{so5ex}
\eea
where we have chosen a real representation of gamma matrices.
With this choice, the first equation of (\ref{conaction})
is the well known magic identity for $SO(5,5)$ gamma-matrices.
The second equation reduces to a non-trivial
$SO(5,5)$ gamma-matrix identity if in addition one imposes
the tensor $\tg^{Ir}$ to be gamma-traceless according to
\bea
g^{Ir} \gamma_{Irs}&=&0\;,
\label{gammatrace}
\eea
i.e.\ to parametrize the real ${\bf 144}_c$ representation.
Some further calculation shows that the remaining equations of (\ref{conaction})
which are quadratic in $\tg^{Ir}$ then reduce
to the last two equations which transform in the ${\bf 10}+{\bf 126_c}+{\bf 320}$ of
$SO(5,5)$\,. A particular solution to these equations can be found
by choosing $g^{Ir} $ to live within the ${\bf 15}\subset {\bf 144_c}$
upon breaking to the maximal subgroup $GL(5)\subset SO(5,5)$\,.
This simply follows from the fact that the symmetric tensor product $({\bf 15} \otimes {\bf 15})_{\rm sym}$
does not contain any representation that lies in the ${\bf 10}+{\bf 126_c}+{\bf 320}$
in which the bilinear constraints transform. Representing the 15 parameters as a symmetric
$5\times 5$ matrix, the resulting gauge group is $CSO(p,q,r)$ with $p+q+r=5$ according
to the signature of the matrix, cf.~\cite{Bergshoeff:2007ef} for details.
In particular, these gaugings include the theory with compact gauge group $SO(5)$.
It is instructive, to give the bosonic field content in representations of this gauge group:
\bea
A_\mu^r &\longrightarrow& {\bf 1}_{+5} + {\bf 5}_{-3} + {\bf 10}_{+1}
\;,
\nonumber\\
(\phi^I , B^I_{\mu\nu}) &\longrightarrow& {\bf 5}_{+2} + {\bf 5}_{-2}
\;,
\nonumber\\
C_{\mu\nu\rho\,r} &\longrightarrow& {\bf 1}_{-5} + {\bf 5}_{+3} + {\bf 10}_{-1}
\;,
\label{repsSO55}
\eea
where the subscripts refer to $GL(1)$ charges under the embedding
$GL(1)\times SO(5) \subset GL(5) \subset SO(5,5)$,
under which the tensor $g^{Ir}$ has charge $-1$.
In particular, the tensor multiplets transform in two copies of the fundamental representation
of the gauge group.
The scalar field content shows that the gauge invariant cubic potential of
(\ref{action}) for this theory vanishes identically.

In order to elucidate the structure of the $SO(5)$ theory, we will calculate the fluctuations
around a maximally supersymmetric solution
according to general analysis of section~\ref{subsec:ex}.
It follows from (\ref{vacuum0}) and the particular form of $\tg^{Ir}$ that maximal supersymmetry
of the vacuum amounts to restricting $\phi^I_0$
to values within the ${\bf 5}_{+2}$ of (\ref{repsSO55}).
For the supersymmetric $SO(5)$ invariant vacuum $\phi^I_0=0$,
both matrices $K_{rs}$, $N_{Ir}$ from (\ref{KN}) vanish identically. As a result, the linearized field
equations (\ref{eom_linearized_action}) simply
describe ten copies of the self-dual tensor multiplet whereas as discussed above
in the unbroken phase, the vector multiplets are invisible
in this perturbative analysis. In the notation of section~\ref{subsec:ex} we find five copies of (T) and
of (T'), respectively.

Let us instead consider a non-vanishing value of $\phi^I_0$ in the ${\bf 5}_{+2}$ which breaks the
gauge symmetry at the vacuum down to $SO(4)$ but preserves all supersymmetries.
Accordingly, the bosonic fields break into
\bea
A_\mu^r &\longrightarrow& {\bf 1}_{+5} + {\bf 1}_{-3} + {\bf 4}_{-3} + {\bf 4}_{+1} + {\bf 6}_{+1}
\;,
\nonumber\\
(\phi^I , B^I_{\mu\nu}) &\longrightarrow& {\bf 1}_{+2} + {\bf 4}_{+2}+ {\bf 1}_{-2} + {\bf 4}_{-2}
\;,
\nonumber\\
C_{\mu\nu\rho\,r} &\longrightarrow& {\bf 1}_{-5} + {\bf 1}_{+3} + {\bf 4}_{+3} + {\bf 4}_{-1} + {\bf 6}_{-1}
\;,
\label{repsSO4}
\eea
under $SO(4)\times GL(1)$\,.
In this case, the only non-vanishing entries in the kinetic vector matrix
$K_{rs}$ are the off-diagonal entries in its
${\bf 4}_{+3} \times {\bf 4}_{-1}$ and ${\bf 4}_{-1} \times {\bf 4}_{+3}$ blocks,
corresponding to eight non-vanishing eigenvalues,
of which four are negative.
Accordingly, the vector fields from the ${\bf 1}_{+5}+{\bf 1}_{-3}+{\bf 6}_{+1}$
(which include the fields in the adjoint representation of the unbroken gauge group)
do not appear in the lowest order fluctuations~(\ref{eom_linearized_action}).
On the other hand, the matrix $\tg^{Ir}$ as chosen above
has its only non-vanishing entries in the
$({\bf 1}_{+2} + {\bf 4}_{+2})\times({\bf 1}_{-3} + {\bf 4}_{-3})$ block.
This shows in particular, that from the three-form fields $C_{\mu\nu\rho\,r}$,
only the components
in the ${\bf 1}_{+3} + {\bf 4}_{+3}$ appear in the action (\ref{action}).
Evaluating the linearized field equations
(\ref{eom_linearized_action}) for these fields,
one verifies that these indeed fall into the structures
identified in (\ref{excitations}).
The explicit result for the representation content of
the various multiplets is displayed in table~\ref{tab:so5}.
In order to correctly keep track of the $GL(1)$ charges,
it is worth to keep in mind that
the gauge coupling constant $g$ and the
scalar vacuum
expectation value $\kappa$
appearing in these equations are of charge $-1$ and $+2$, respectively.

\begin{table}[bt]
\begin{center}
\begin{tabular}{|c|c|c|c|}
\hline
 (T) & (T') & (TV) & (VT)\\ \hline \hline
${\bf 1}_{-2}$: $\!(\varphi^-, \chi^-, B^-)\!$ & ${\bf 1}_{+3}$: $C^+$
&${\bf 4}_{+3}$: $C^m$&
\\
& ${\bf 1}_{+2}$: $\!(\varphi^+, \chi^+, B^+)\!$ &  ${\bf 4}_{+2}$: $\!(\varphi^m, \chi^m, B^m)\!$ &
${\bf 4}_{-2}$ $\!(\tilde\varphi^m, \tilde\chi^m, \tilde B^m)\!$
\\
&&${\bf 4}_{+1}$: $\!(A^m, \lambda^m)\!$&${\bf 4}_{-3}$: $\!(\tilde A^m, \tilde \lambda^m)\!$
\\ \hline
\end{tabular}
\caption{Lowest order fluctuations around the $SO(4)$
invariant vacuum.}
\label{tab:so5}
\end{center}
\end{table}

\subsection{Example: Nilpotent gauge group}
\label{subsec:solutions2}

Another solution to the constraints (\ref{conaction})
may be obtained from the gauged supergravities of~\cite{Gunaydin:2010fi}.
In this case, vector and tensor multiplets are supposed to come in the
spinor and vector representation, respectively, of the group $SO(9,1)$.
Since real gamma matrices exist, and their
algebra is the same as in the previous example,
with the choice (\ref{so5ex}), the first two equations of (\ref{conaction})
again reduce to gamma-tracelessness (\ref{gammatrace}) of the tensor $\tg^{Ir}$\,.
However, in this case, the remaining constraint equations turn out to admit
a unique solution, which is given by
\bea
\tg^{Ir} &\equiv& g\, \zeta^r \zeta^s \zeta^t \gamma^I_{st}
\;,
\label{tzzz}
\eea
with gauge coupling constant $g$ and
an arbitrary constant $SO(9,1)$ spinor $\zeta^r$. This choice corresponds to a nilpotent gauge group
whose algebra $N^+_{8}$ is embedded into $\mathfrak{so}(9,1)$ according to the three-grading
\bea
\mathfrak{so}(9,1)&\longrightarrow&
N^-_{8} \oplus \left(\mathfrak{so}(8)\oplus \mathfrak{so}(1,1)\right) \oplus N^+_{8}
\;,
\label{Ntranslations}
\eea
see~\cite{Gunaydin:2010fi} for further details.
Under the little group $SO(7)$ of the spinor defining (\ref{tzzz}), the multiplets decompose as
\bea
A_\mu^r &\longrightarrow& {\bf 1}_{-1} + {\bf 7}_{-1} + {\bf 8}_{+1}
\;,
\nonumber\\
(\phi^I , B^I_{\mu\nu}) &\longrightarrow& {\bf 1}_{+2} + {\bf 1}_{-2} + {\bf 8}_{0}
\;,
\nonumber\\
C_{\mu\nu\rho\,r} &\longrightarrow& {\bf 1}_{+1} + {\bf 7}_{+1} + {\bf 8}_{-1}
\;,
\label{repsSO91}
\eea
where again we keep the charges under the $GL(1)$ under which the
gauge coupling constant carries charge $-3$.
A distinctive feature of this model as compared to the previous one, is a nonvanishing
cubic scalar potential. More precisely, the scalar Lagrangian takes the form
\bea
{\cal L} &=&
-\ft18 D^\mu \phi^i \,D_\mu \phi^i -\ft18 \partial^\mu \phi^+ D_\mu \phi^-
+ g^3 (\phi^+)^3
\;,
\eea
where $(\phi^+,\phi^-,\phi^i)$ represent the ${\bf 1}_{+2} + {\bf 1}_{-2} + {\bf 8}_{0}$
scalars in the ${\bf 8}_0$ according to the decomposition (\ref{repsSO91}).
A maximally supersymmetric vacuum is found by choosing a non-vanishing $\phi_0^i$,
which breaks one generator of the nilpotent gauge group, and
the little group down from $SO(7)$ to $G_2$.
In this case, the matrix $K_{rs}$ in (\ref{eom_linearized_action}) remains invertible,
such that all fields contribute to the linearized fluctuation equations.
Evaluating the fluctuation equations, one
confirms that all fluctuations again
fall into the structures
identified in (\ref{excitations}).
The final result for the representation content of
the various multiplets is displayed in table~\ref{tab:so9}.

\begin{table}[bt]
\begin{center}
\begin{tabular}{|c|c|c|c|}
\hline
(V) & (T) &  (TV) & (VT)\\ \hline \hline
${\bf 7}+{\bf 7}$
&
${\bf 7}+{\bf 1}$
&
${\bf 1}$
&
${\bf 1}$
\\ \hline
\end{tabular}
\caption{Lowest order fluctuations around the $G_2$
invariant vacuum.}
\label{tab:so9}
\end{center}
\end{table}

\section{Conclusions}
\label{sec:conclusions}

In this paper, we have constructed
a general class of six-dimensional (1,0) superconformal models with
non-abelian vector and tensor multiplets.
The construction is based on the non-abelian hierarchy of $p$-form fields
and strongly relies on the introduction of further three-form gauge potentials.
These are related to the vector fields by a first-order duality equation
and do not constitute new degrees of freedom, however they play a crucial
role in communicating the degrees of freedom between the vector and tensor multiplets.
The models are parametrized by a set of dimensionless constant tensors,
which are constrained to satisfy a number of algebraic identities (\ref{lincon})--(\ref{quadcon}).
Generically these models provide only equations of motions which we have derived
from closure of the supersymmetry algebra. For particular choice of the parameters,
the equations of motion may be integrated to an action. However, the
kinetic metrics in the vector and the tensor sector appear with indefinite signature.
It will require further work to understand the fate of the resulting ghost states
and if one can for example decoupled them
with the help of the large extended tensor gauge symmetry.
For the M2-brane theories, a similar structure has appeared
in the theories based on Lorentzian signature 3-algebras. In these models,
the ghost states have been eliminated at the cost of breaking
conformal symmetry
by further gauging of particular shift symmetries~\cite{Bandres:2008kj},
which are however absent in the models constructed here.
The cubic potential, if non-vanishing,  will generically be unbounded from below.
However, since the indefinite metric brings in negative norm states,
the relation $E = || Q ||^2 \geq 0$ is no longer valid in such cases
($E$ = energy, $Q$ = supercharge) and a non-vanishing cubic potential is
in principle possible and not forbidden by supersymmetry.

We have discussed several explicit examples which satisfy all algebraic consistency conditions.
An arbitrary compact gauge group with tensor fields in the adjoint representation can be realized
on the level
of equations of motion. Lagrangian models have been given for the
compact gauge group $SO(5)$ and for a particular eight-dimensional
nilpotent gauge group embedded in $SO(9,1)$. All these models share
some peculiar features. The fluctuation spectrum of excitations around a supersymmetric
vacuum contain not only free vector and tensor multiplets, but also certain
`non-decomposable' combinations of couplings between the two,
which we have collected in (\ref{excitations}).
Moreover, null-directions in the kinetic vector matrix may appear for unbroken gauge symmetries
and cause that the fields of the corresponding vector multiplets drop out of this perturbative analysis.
In general this analysis  is valid only in the spontaneously broken phase,
however, the unbroken sector of the Yang-Mills multiplet is still
(infinitely) strongly coupled and perturbatively not visible.
The corresponding part of the spectrum decouples and should be
integrated out for a proper treatment.

Let us note that although we have used in our explicit examples the algebraic structure
underlying gauged supergravity theories in order to find solutions to the algebraic
consistency constraints (\ref{conaction}), none of these theories can be obtained as a
suitable flat-space limit of the supergravities
of~\cite{Duff:1997me, Bergshoeff:2007ef,Gunaydin:2010fi}.
E.g.\ globally supersymmetric theories derived as a flat space limit of these supergravities
(if they exist) would not have ghosts in the scalar sector, which seems to be an inevitable
feature of the theories presented here.

An obvious direction of further investigation is the study of the constraints
(\ref{conaction}) and (\ref{lincon})--(\ref{quadcon}) for models with and without
action, respectively. Especially for the case with an action, it would be
highly interesting to understand, if the model with compact $SO(5)$ gauge group
that we have presented in section~\ref{subsec:solutions1} corresponds to very particular
solution of these constraints or if it may be generalized to other gauge groups.
In this context, it may be interesting to pursue the comparison to the five-dimensional
superconformal models classified and studied in~\cite{Bergshoeff:2002qk},
which may elucidate the geometric role of the set of algebraic
consistency constraints (\ref{conaction}) that underlie our construction.
Another interesting research direction is the generalization of
 the analysis our maximally supersymmetric vacua of these models
to such states which only preserve a fraction of supersymmetry.

An intriguing question about the models is, how much of the presented structures can be carried
over to $(2,0)$ theories. Although there is no propagating-$(2,0)$ vector multiplet, the present construction
has illustrated the possible relevance of the inclusion of non-propagating degrees of freedom.
As a first step in this direction, we have briefly sketched in section~\ref{subsec:hypers}
the inclusion of hypermultiplets to the gauged models. Adding $n_{\rm T}$ hypermultiplets
with flat target space completes the present field content from (1,0) to the (2,0) theories.
A different extension of our models within the (1,0) framework
could be obtained by studying the possibilities of coupling linear multiplets
as sketched in~\cite{Sezgin:1994th}.
A pending question is of course the quantization of the models, in particular
the decoupling of the ostensible ghost states and if the conformal symmetry is
preserved at the quantum level.
Last but not least, the study of anomalies for the presented models with
their new gauge symmetries and non-abelian couplings raises an entirely new set of questions.

It seems clear from our discussion and the many open questions
that we are still far from a profound understanding
of the models we have presented in this paper.
On the other hand, given the hitherto lack of non-abelian models in six dimensions
the very existence of these models is rather fascinating.
They provide new and very intriguing structures that deserve more study
and may yet reserve further surprises.
We look forward to further analysis.

\subsection*{Acknowledgement}

This work is supported in part by the Agence Nationale de la Recherche (ANR).
We wish to thank F.\ Delduc for helpful discussions. We are grateful
to Linus Wulff for his help in the construction of the note added in Section 3.6.
R.W.\ thanks Yu-tin Huang for useful discussions in the early stages
of the project and E.S. thanks University of Lyon for hospitality. The research
of E.~S. is supported in part by NSF grants PHY-0555575 and PHY-0906222.
\bigskip

\begin{appendix}

\section*{Appendix}

\section{Conventions}

In this appendix, we summarize our space-time and spinor conventions.
We work with a flat space-time metric of signature $(-+++++)$
and Levi-Civita tensor $\varepsilon_{012345}=1$.
(Anti-)selfdual three-forms are defined such as to satisfy
\bea
H_{\mu\nu\rho}^{\pm} &=&
\pm \frac{1}{3!}\,\varepsilon_{\lambda\sigma\tau\mu\nu\rho}\,H^{\lambda\sigma\tau\,\pm}
\;,\label{Hplus}
\eea
with $H_{\mu\nu\rho}=H_{\mu\nu\rho}^++H_{\mu\nu\rho}^-$\,.
Six-dimensional gamma-matrices satisfy the basic relations
\bea
\{\gamma_\mu , \gamma_\nu \} &=& 2 \tg_{\mu\nu}\;,\nonumber\\
\gamma_7 &\equiv& \gamma_{012345}\;,\qquad \gamma_7^2 ~=~ 1
\;,\nonumber\\
\gamma^{a_1 \cdots a_n} &=& \frac{s_n}{(6-n)!}\,
\varepsilon^{a_1 \cdots a_n b_1 \cdots b_{6-n}} \gamma_{b_1 \cdots b_{6-n}} \, \gamma_7
\;,\qquad
s_n = \left\{
\begin{array}{rl}
-1: & n=0,1,4,5\\
+1: & n=2,3,6
\end{array}
\right.
\;,
\nonumber\\
\eea
as well as the particular identities
$\gamma^\lambda \gamma_{\mu\nu\rho} \gamma_\lambda=0$,
$\gamma^{\mu\nu\rho} \gamma_{\lambda} \gamma_{\mu\nu\rho}=0$,
$\gamma^{\mu\nu\rho} \gamma_{\lambda\sigma\tau} \gamma_{\mu\nu\rho}=0$.
\medskip

The spinor chiralities are given by
Spinor chiralities
\bea
\gamma_7 \,\epsilon = \epsilon\;,\quad
\gamma_7\, \lambda^r =  \lambda^r\;,\quad
\gamma_7 \,\chi^X = -\chi^X
\;.
\eea
In addition, the fermions carry $Sp(1)$ indices for which we use
standard northwest-southeast conventions according to
$\lambda^i = \varepsilon^{ij} \lambda_j$, etc.
Accordingly, their bilinear products satisfy the
symmetry properties
\bea
\bar{\lambda}^i \gamma^{(n)} \chi^j &=& t_n \bar{\chi}^j \gamma^{(n)} \lambda^i
\;,\qquad
t_n = \left\{
\begin{array}{rl}
-1: & n=0,3,4\\
+1: & n=1,2,5,6
\end{array}
\right.
\;.
\eea
The Fierz identities are of the form
\bea
\epsilon_2^j \bar{\epsilon}_1^i &=&
-\frac1{4}\, \left(\xi^\mu \gamma_\mu \varepsilon^{ij} +\frac1{6}\,\xi^{ij}_{\mu\nu\rho}\,\gamma^{\mu\nu\rho}\right)
\frac{1-\gamma_7}{2}\;,
\nonumber\\[2ex]
&&\mbox{with}\quad
\xi^\mu \equiv \ft12\bar{\epsilon_2} \gamma^\mu \epsilon_1
\;,
\quad
\xi^{ij}_{\mu\nu\rho} \equiv \ft12\bar{\epsilon_2}^i \gamma_{\mu\nu\rho} \epsilon_1^j
\;.
\eea
In addition, we will employ some particular Fierz identities, cubic
in a spinor~$\lambda^r$
\bea
0&=&
t_{rs,uv}
\left(
3\,\bar{\epsilon} \gamma^\rho \lambda^u \,\bar\lambda{}^s \gamma_{\mu\nu\rho}\lambda^v
+4\,\bar{\epsilon} \gamma_{[\mu} \lambda^u \,\bar\lambda{}^s \gamma_{\nu]}\lambda^v
-\bar{\epsilon} \gamma_{\mu\nu\rho} \lambda^u \,\bar\lambda{}^s  \gamma^\rho \lambda^v
\right)
\;,\label{id1}\\[1ex]
0&=&
t_{rs,uv}
\left(
\bar{\epsilon}_{(i} \gamma^\mu \lambda_{j)}^u \,\bar\lambda{}^s \gamma_{\mu}\lambda^v
-3\,\bar{\epsilon} \gamma_\mu \lambda^u \,\bar\lambda_{(i}^s \gamma^{\mu}\lambda_{j)}^v
\right)
\;,
\label{id2}
\eea
with an arbitrary tensor $t_{rs,uv}=t_{[rs],(uv)}$ satisfying $t_{r(s,uv)}=0$\,.
These identities can be derived by making use of the following well known $\gamma$-matrix identity
\be
\eta_{\mu\nu}\, \gamma^\mu_{\delta\ell,(\alpha i}\,\gamma^\nu_{\beta j,\gamma k)}=0 \label{magic}
\ee
as follows. Multiplication of this identity by
\be
t_{rs,uv} \left(\gamma^{\rho\sigma}\right)_{\eta m}{}^{\alpha i} \,\epsilon^{\delta\ell} \lambda^{\beta j,u} \,\lambda^{\gamma k,s}\,\lambda^{\eta m,v}
\ee
and using the conventions
\be
\gamma^\mu_{\alpha i,\beta j} = \gamma_{\alpha\beta} \varepsilon_{ij}\ ,\qquad \lambda^{\alpha i}\gamma^\mu_{\alpha\beta} \lambda^\beta_j = {\bar\lambda}^i\gamma^\mu \lambda_j\
\ee
yields the identity \eqn{id1}. Similarly, multiplication of \eqn{magic} with
\be
t_{rs,uv} \lambda^{\delta n,s}\,\lambda^{\alpha u}_j\, \lambda^{\beta k,v}\,\epsilon^{\gamma \ell}
\ee
produces the identity \eqn{id2}.

\end{appendix}

\newpage


\bibliographystyle{JHEP2}

\providecommand{\href}[2]{#2}\begingroup\raggedright\endgroup

\end{document}